\documentstyle[12pt]{article}
\begin{document}
\begin{flushright}
hep-th/0403230\\
SNB/March/2004
\end{flushright}
\vskip 2cm
\begin{center}
{\bf \Large {Geometrical Aspects of BRST Cohomology 
\\in Augmented Superfield Formalism }}

\vskip 2.5cm

{\bf R.P.Malik}
\footnote{ E-mail address: malik@boson.bose.res.in  }\\
{\it S. N. Bose National Centre for Basic Sciences,} \\
{\it Block-JD, Sector-III, Salt Lake, Calcutta-700 098, India} \\

\vskip 2cm

\end{center}

\noindent
{\bf Abstract}:
In the framework of augmented superfield approach, we provide the 
geometrical origin and interpretation for the nilpotent (anti-)BRST charges, 
(anti-)co-BRST charges and a non-nilpotent bosonic charge. Together, these
local and conserved charges  turn out to be responsible
for a clear and cogent definition of the Hodge decomposition theorem
in the quantum Hilbert space of states. The above charges owe their origin
to the de Rham cohomological operators of differential geometry which
are found to be at the heart of some of the key concepts associated with 
the interacting gauge theories. For our present review, we choose the 
two $(1 + 1)$-dimensional (2D)
quantum electrodynamics (QED) as a prototype field 
theoretical model to derive all the nilpotent symmetries
for  all the fields present in this interacting gauge theory in the
framework of augmented superfield formulation and show that this theory
is a {\it unique} example of an interacting gauge theory which provides a
tractable field theoretical model for the Hodge theory. 

\baselineskip=16pt

\vskip 1cm



\newpage

\noindent
{\bf 1 Introduction}\\

\noindent
One of the most elegant and intuitive approaches to covariantly quantize
a gauge theory, endowed with the first-class constraints [1]
in the language of the Dirac's classification scheme, is the 
Becchi-Rouet-Stora-Tyutin (BRST) formalism 
(see, e.g, [2]). In this formalism, the unitarity
and ``quantum'' gauge (i.e. BRST) invariance are respected together at any
arbitrary order of perturbative computations connected with the physical 
processes involving the gauge fields and the matter fields of any arbitrary
interacting 1-form (non-)Abelian gauge theories (see, e.g. [3,4])
\footnote{The full strength of the BRST formalism turns up
in its full glory in the context of
non-Abelian gauge theories where, for each loop diagram consisting of
the gauge (gluon) fields {\it alone}, there exists a diagram consisting of the 
(anti-)ghost fields as its counterpart
so that the unitarity in the theory can be maintained
at any arbitrary order of perturbative computation
(see, e.g., [3] for detailed computations).}. The BRST formalism
is indispensable in the context of 
the modern developments in the topological field
theories and topological string theories [5-7]. The range of the applicability 
of this formalism has been extended by the inclusion of the second-class 
constraints in its ever widening framework (see, e.g., [8]). Its deep
connections with the mathematics of cohomology and differential geometry,
its beautiful interpretation in the framework of superfield formulation,
its intimate relationships with the basic tenets of supersymmetry, its
fruitful application in the context of reparametrization invariant
theories such as supergravity theories, superstring theories, brane
dynamics, M-theory, etc., have elevated this area of research to a high
degree of mathematical sophistication as well as physical applications
(see, e.g., [9,10] and references therein).

The geometrical aspects of any arbitrary physical theory provide
a clear physical interpretation for 
the key theoretical ideas expressed in their abstract 
mathematical form. One of the most attractive, accurate 
and intuitive theoretical approaches to gain
an insight into the {\it geometrical} origin of the existence of the nilpotent
(anti-)BRST symmetry transformations (and their corresponding conserved
and nilpotent generators), for any arbitrary
 $p$-form (non-)Abelian gauge theories, is the
superfield formalism [11-18]. In this formalism, a $(p + 1)$-form 
super curvature $\tilde F = \tilde d \tilde A + \tilde A \wedge \tilde A$
is constructed from the super exterior derivative 
$\tilde d = dx^\mu \partial_\mu + d \theta \partial_\theta + d \bar\theta
\partial_{\bar\theta}$ (with $\tilde d^2 = 0)$ and a super $p$-form 
($p = 1, 2, 3, .....)$ connection $\tilde A$ on a $(D + 2)$-dimensional
supermanifold parametrized by the superspace variables $Z^M = (x^\mu,
\theta, \bar\theta)$ where $x^\mu (\mu = 0, 1, 2......D - 1)$ are the
$D$-number of even (bosonic) spacetime coordinates and $\theta$ and
$\bar\theta$ are a pair of Grassmannian (odd) variables (i.e.
$\theta^2 = \bar\theta^2 = 0, \theta \bar\theta + \bar\theta \theta = 0$).
This $(p + 1)$-form super curvature is subsequently equated, 
due to the so-called
horizontality condition
\footnote{This condition is referred to as the soul-flatness condition
by Nakanishi and Ojima [19].}, with the
$(p + 1)$-form ordinary curvature $ F = d A + A \wedge A$ defined on
the ordinary $D$-dimensional flat Minkowski manifold with the help of
an ordinary exterior derivative $d = dx^\mu \partial_\mu$
(with $d^2 = 0$) and a $p$-form ordinary connection $A $. 
This condition leads to the derivation of the
nilpotent (anti-)BRST symmetry transformations for the gauge field and the
(anti-)ghost fields of the $p$-form gauge theory. We christen this superfield
formulation as the {\it usual} superfield approach because, as is evident, this
formalism does not shed any light on the derivation of the nilpotent
(anti-)BRST symmetry transformations for the matter fields of an
interacting gauge theory
(where there is a coupling between the gauge field and matter fields). 
However, this approach does provide the geometrical
origin and interpretation for (i) the (anti-)BRST symmetry transformations
and their corresponding generators as the translation generators 
($ \mbox{Lim}_{\bar\theta \to 0}(\partial/\partial\theta), 
\mbox{Lim}_{\theta \to 0} (\partial/\partial\bar\theta)$)
along the Grassmannian directions of
the $(D + 2)$-dimensional supermanifold, (ii) the nilpotency property
associated with the (anti-)BRST symmetry transformations (and corresponding 
conserved and nilpotent generators) as a couple of successive translations
(i.e. $(\partial/\partial \theta)^2 = (\partial/\partial\bar\theta)^2 = 0$)
along either the $\theta$- or 
$\bar\theta$-directions of the supermanifold, and
(iii) the anti-commutativity property of the local, covariant, continuous
and nilpotent
BRST and anti-BRST symmetry
transformations (and corresponding 
conserved and nilpotent (anti-)BRST charges) as a couple
of translations along the $(\theta)$- and $(\bar\theta)$-directions followed
by another couple of translations along the Grassmannian directions
with the reverse order
(i.e. $(\partial/\partial\theta) (\partial/\partial\bar\theta)
+ (\partial/\partial\bar\theta) (\partial/\partial\theta) = 0$).

It is obvious from the above discussions that only one (i.e. $\tilde d$)
of the three
\footnote{ On an ordinary manifold without a boundary, the set of
operators $(d, \delta, \Delta)$ define the de Rham cohomological
operators of the differential geometry. The (co-)exterior derivatives 
$(\delta)d$ and the Laplacian
operator $\Delta = d \delta + \delta d$ obey an algebra:
$d^2 = \delta^2 = 0, \Delta = (d + \delta)^2, [\Delta, d] = 0,
[\Delta, \delta] = 0$ where $ d = dx^\mu \partial_\mu$ and
$\delta = \pm * d *$ are connected with each-other by the Hodge duality
($*$) operation defined on the manifold  and the $(+)-$ signs
(present in $\delta = \pm * d *$) depend on 
(i) the degree of the differential forms 
involved in the inner product, and (ii) the dimensionality of 
the manifold. The Hodge decomposition 
(i.e. $f_n = \omega_n + d e_{n -1} + \delta c_{n+1}$) of an arbitrary
differential form $f_n$ (of degree $n$) owes its origin to
the above cohomological operators that define $\omega_n$ as the harmonic form
(i.e. $d \omega_n = \delta \omega_n = 0$),
$d e_{n-1}$ as the exact form and $\delta c_{n+1}$ as the co-exact form
into which the arbitrary form $f_n$ is decomposed on the manifold
(see, e.g. [20-24] for details).} 
 super de Rham cohomological operators 
$(\tilde d, \tilde \delta, \tilde \Delta$) has been exploited in
the horizontality condition that provides the geometrical interpretation for 
the conserved and nilpotent (anti-)BRST charges which generate the local,
covariant, continuous and nilpotent symmetries for the gauge theories. 
In the usual discussions on the BRST cohomology, only the nilpotent
BRST charge (i.e. the analogue of the cohomological operator $d$) plays
the central role. However, a thorough discussion on the BRST cohomology
requires the existence of a nilpotent co-BRST charge (i.e. the analogue
of the cohomological operator $\delta = \pm * d *$) so that the
celebrated Hodge decomposition theorem can be defined precisely. Towards
accomplishing this goal, in a 
recent set of papers [25-28], all the super de Rham cohomological operators
have been exploited in the framework of the {\it usual} superfield
formulation to derive the local, covariant
and continuous (anti-)BRST, (anti-)co-BRST
and a bosonic symmetry transformations for the gauge- and (anti-)ghost fields
of the two ($1 + 1$)-dimensional free (non-)Abelian gauge theories defined
on a four $(2 + 2)$-dimensional supermanifold. In these derivations,
the generalized versions of the horizontality condition play
 very decisive roles. The geometrical
interpretations for all the above symmetries have been provided in
the framework of the {\it usual}
superfield formulation. Interestingly, for the first time,
the geometrical meaning of the topological nature of the above field theories
has been provided in the framework of {\it usual}
superfield formulation where it has been shown that the Lagrangian density 
and the symmetric energy momentum tensor for these theories
 can be interpreted as the translation of some local (but 
composite) superfields along the Grassmannian directions of the supermanifold
[29,30]. However, in the above attempts [25-30] too, there is no discussion 
on the derivations of the above symmetry transformations for the matter fields 
present in an {\it interacting} gauge theory
(which is more natural than the free gauge theories). Thus, the results of the 
{\it usual}
superfield approach [11-18] (as well as its extended version
[25-30]) are still incomplete as far as the derivation of {\it all}
the nilpotent transformations for {\it all} the fields 
present in an interacting gauge theory is concerned.

In our recent set of papers [31-34], the above superfield formulations, 
endowed with the generalized versions of the horizontality condition, have
been augmented by the additional restrictions that owe their origin to the
requirement of the invariance of the matter conserved currents
on the (super-)manifolds. We christen this version of
the superfield approach as the  augmented superfield formalism because it
sheds light on the derivation of the nilpotent symmetries for {\it all}
the fields (including the matter fields)
present in an interacting gauge theory. The purpose of the present
paper is to derive all the off-shell as well as the on-shell nilpotent 
(anti-)BRST and (anti-)co-BRST symmetry transformations for {\it all}
the fields present in the BRST invariant Lagrangian density
(cf. (2.1),(2.2) below) for the two $(1 + 1)$-dimensional
interacting gauge theory (i.e. 2D QED
\footnote{A dynamically closed system of the photon field and
the Dirac fields in 2D.}) in the framework of augmented
superfield formalism. We show that the off-shell nilpotent (anti-)BRST
as well as (anti-)co-BRST symmetry transformations (and their corresponding
nilpotent generators) correspond to the translation generators
($\mbox{Lim}_{\bar\theta \to 0} (\partial/\partial\theta),
\mbox{Lim}_{\theta \to 0} (\partial/\partial\bar\theta)$) along the
Grassmannian $(\theta)\bar\theta$-directions of the 
four $(2 + 2)$-dimensional supermanifold. The process
of translation of the general superfields produces the internal (anti-)BRST
and (anti-)co-BRST symmetry transformations on their counterparts
ordinary fields, defined on the ordinary Minkowski flat manifold. We also
demonstrate that the on-shell nilpotent (anti-)BRST and (anti-)co-BRST
symmetry transformations owe their origin to (anti-)chiral superfields
where the on-shell nilpotent conserved charges (and their corresponding
symmetries) are interpreted as the translation generators
($(\partial/\partial\theta), (\partial/\partial\bar\theta)$) along a
specific Grassmannian direction (i.e. $\theta$ or $\bar\theta$) of
the three $(2 + 1)$-dimensional (anti-)chiral super sub-manifold of 
the most general four $(2 + 2)$-dimensional supermanifold. We have chosen, for
our discussions, the 2D QED as a prototype  example of 
an interacting gauge theory 
primarily for three reasons. First, the dual(co-)BRST transformations
for this theory are local, covariant, continuous and nilpotent [35,36]. This is
in sheer contrast with the interacting 4D Abelian gauge theory where
the corresponding transformations are non-local, non-covariant, continuous
and nilpotent (see, e.g., [37,38]). Second, it is a {\it unique}
{\it interacting} gauge theory which provides a tractable field theoretical
model for the Hodge theory (see, e.g., [36]). The other set of
field theoretical
models, that have been shown to be the examples of the Hodge theory,
are (i) the free 2D Abelian 1-from 
gauge theory [39-41],
(ii) the self-interacting 2D non-Abelian 1-form gauge theory (without any
interaction with matter fields) [41,42], and (iii) the 4D free Abelian 2-form
gauge theory [43,44]. It should be noted that there exist no interaction terms
(involving the gauge field and matter fields)
in the field theoretical models cited in (i), (ii) and (iii). Finally, 2D QED
is a field theoretical model where the {\it topological} gauge field
$A_\mu$ (see, e.g., [41] for details)
couples with the matter conserved current $J_\mu \sim \bar\psi
\gamma_\mu \psi$. Thus, 2D QED is interesting by itself because it is
a very specific model for the
{\it interacting} topological field theory [35,36].

The material of our present paper is organized as follows. In Section 2, we 
recapitulate the bare essentials of our earlier investigations [35,36] 
on the existence of the
on-shell as well as off-shell nilpotent symmetries, a bosonic symmetry and
a ghost (scale) symmetry transformations for the 2D interacting gauge
theory (i.e. QED) in the framework of Lagrangian formulation. Section 3
is devoted to the derivation of the on-shell as well off-shell nilpotent
symmetries for the gauge- and (anti-)ghost fields in the framework of
{\it usual} superfield formulation. In Section 4, we deal with a detailed
discussion (and an accurate derivation) of the nilpotent symmetries for the 
matter (Dirac) fields of this interacting gauge theory in the framework
of augmented superfield formalism. Finally, in Section 5, we make some
concluding remarks and indicate briefly the future perspective of
our present endeavour.\\

\noindent
{\bf 2 Preliminary: (Anti-)BRST- and (Anti-)co-BRST Symmetries}\\

\noindent
We briefly demonstrate here the existence of the nilpotent (anti-)BRST
and (anti-)co-BRST symmetry transformations for the interacting two
$(1 + 1)$-dimensional (2D) QED in the framework of Lagrangian formulation [36].
To this end in mind, we begin with the following Lagrangian density
for the above interacting physical system in the Feynman gauge
\footnote{ We adopt here the conventions and notations such that
the flat 2D Minkowski metric is $\eta_{\mu\nu} = $ diag $(+1, -1)$ and
$\Box = \eta^{\mu\nu} \partial_\mu \partial_\nu = (\partial_0)^2 -
(\partial_1)^2, \varepsilon_{01} = \varepsilon^{10} = + 1,
F_{01} = - \varepsilon^{\mu\nu} \partial_\mu A_\nu = F^{10}
= \partial_o A_1 - \partial_1 A_0, D_\mu \psi = \partial_\mu \psi 
+ i e A_\mu \psi$. The Dirac matrices in the two-dimensional spacetime
are: $\gamma^0 = \sigma_2, \gamma^1 = i \sigma_1, \gamma_5 = 
\gamma^0 \gamma^1 = \sigma_3, \{\gamma^\mu, \gamma^\nu \} = 2 \eta^{\mu\nu},
\gamma_\mu \gamma_5 = \varepsilon_{\mu\nu} \gamma^\nu$. Here the Greek indices
$\mu, \nu, \rho .......= 0, 1$ stand for the spacetime directions on the
the 2D Minkowski manifold and $\sigma$'s are the usual $2 \times 2$
Pauli matrices.} [4,19,45,46]
$$
\begin{array}{lcl}
{\cal L}_{b} &=& - \frac{1}{4}\; F^{\mu\nu} F_{\mu\nu} 
+ \bar \psi \;(i \gamma^\mu D_\mu - m)\; \psi + B \;(\partial \cdot A)
+ \frac{1}{2}\; B^2
- i \;\partial_{\mu} \bar C \partial^\mu C, \nonumber\\
&\equiv& \frac{1}{2}\; E^2
+ \bar \psi\; (i \gamma^\mu D_\mu - m) \;\psi + B \;(\partial \cdot A)
+ \frac{1}{2}\; B^2
- i \;\partial_{\mu} \bar C \partial^\mu C, 
\end{array} \eqno(2.1)
$$
where $F_{\mu\nu} = \partial_\mu A_\nu - \partial_\nu A_\mu$ is the field 
strength tensor for the $U(1)$ gauge theory that is derived from the 2-form
$d A = \frac{1}{2} (dx^\mu \wedge dx^\nu) F_{\mu\nu}$. 
As is evident, the latter
is  constructed by the application of the exterior derivative 
$d = dx^\mu \partial_\mu$ (with $d^2 = 0)$ on the 1-form $A = dx^\mu A_\mu$
(which defines the vector potential $A_\mu$). It will be noted that in 2D,
$F_{\mu\nu}$ has only the electric component (i.e. $F_{01} = E$) and there
is no magnetic component associated with it. The
gauge-fixing term $(\partial \cdot A 
= \partial_\mu A^\mu \equiv \partial_0 A_0 - \partial_1 A_1)$ 
is derived through the operation of the co-exterior derivative $\delta$ 
(with $\delta = - * d *, \delta^2 = 0$) on the
one-form $A$ (i.e. $\delta A = - * d * A = (\partial \cdot A)$)
where $*$ is the Hodge duality operation. 
The operation of the Laplacian operator $\Delta = d \delta + \delta d$
on the 1-form $A = dx^\mu A_\mu$ leads to $\Delta A = dx^\mu \Box A_\mu$.
This, in turn, produces
the equation of motion for the gauge field $A_\mu$, present
in the gauge-fixed Lagrangian density, if we demand that the Laplace
equation $\Delta A = - e J$  (with a 1-form source term  $J = dx^\mu J_\mu$)
to be satisfied. Here $J_\mu = \bar \psi \gamma_\mu \psi$ is constructed 
by the Dirac fields $\psi$ and $\bar\psi$. 
Two important points to be noted, at this juncture, are (i)
the kinetic energy and gauge-fixing terms of the Lagrangian density (2.1) 
owe their origin to the two  (i.e. $d$ and $\delta$) of the three 
de Rham  cohomological operators (i.e. $d, \delta, \Delta$)
of the differential geometry.  (ii) The 
Laplacian operator $\Delta = d \delta + \delta d$ produces
the equation of motion (i.e. $\Box A_\mu = - e J_\mu$)
that results in from the the above gauge-fixed
Lagrangian density. In the above Lagrangian density, the 
fermionic Dirac fields $(\psi, \bar \psi)$, with the mass $m$ and charge $e$, 
couple to the $U(1)$ gauge field $A_\mu$ (i.e. $ - e \bar \psi \gamma^\mu 
A_\mu \psi$) through the conserved current $J_\mu = \bar \psi \gamma_\mu 
\psi$. The anticommuting ($ C \bar C + \bar C C = 0, C^2 = \bar C^2 = 0,
C \psi + \psi C = 0, \bar \psi C + C \bar\psi = 0$, etc.,) 
(anti-)ghost fields $(\bar C)C$ are required to
maintain the unitarity and ``quantum'' gauge (i.e. BRST) invariance together
at any arbitrary order of perturbation theory.
The kinetic energy term $(\;\frac{1}{2} E^2\;)$
of (2.1) can be linearized by invoking an additional
auxiliary field ${\cal B}$
$$
\begin{array}{lcl}
{\cal L}_{B} =  {\cal B} \; E - \frac{1}{2} {\cal B}^2
+ \bar \psi\; (i \gamma^\mu D_\mu - m)\; \psi + B\; (\partial \cdot A)
+ \frac{1}{2}\; B^2
- i\; \partial_{\mu} \bar C \partial^\mu C,
\end{array} \eqno(2.2)
$$
which is the analogue of the Nakanishi-Lautrup auxiliary field $B$ that is
required to linearize the gauge-fixing term $-\frac{1}{2} (\partial\cdot A)^2$
in (2.1).
The above Lagrangian density
(2.2) respects the following off-shell nilpotent
$(s_{(a)b}^2 = 0,  s_{(a)d}^2 = 0)$ (anti-)BRST ($s_{(a)b}$)
\footnote{We follow here the notations and conventions adopted in [46]. 
In fact, in its full glory, a nilpotent ($\delta_{B}^2 = 0$)
BRST transformation $\delta_{B}$ is equivalent to the product of an 
anticommuting ($\eta C = - C \eta, \eta \bar C = - \bar C\eta,
\eta \psi = - \psi \eta, \eta \bar \psi = - \bar \psi \eta$ etc.)
spacetime independent parameter $\eta$ and $s_{b}$ 
(i.e. $\delta_{B} = \eta \; s_{b}$) where $s_{b}^2 = 0$.} 
- and (anti-)dual(co)-BRST ($s_{(a)d}$) symmetry transformations 
(with $s_b s_{ab} + s_{ab} s_b = 0, s_d s_{ad} + s_{ad} s_d = 0$) [36]
$$
\begin{array}{lcl}
s_{b} A_{\mu} &=& \partial_{\mu} C, \qquad 
s_{b} C = 0, \qquad 
s_{b} \bar C = i B,  \qquad s_b \psi = - i e C \psi, \nonumber\\
s_b \bar \psi &=& - i e \bar \psi C,
\qquad s_{b} {\cal B} = 0, \quad  s_{b} B = 0, \quad
\;s_{b} E = 0, \quad s_b (\partial \cdot A) = \Box C, \nonumber\\
s_{ab} A_{\mu} &=& \partial_{\mu} \bar C, \qquad 
s_{ab} \bar C = 0, \qquad 
s_{ab} C = - i B,  \qquad s_{ab} \psi = - i e \bar C \psi, \nonumber\\
s_{ab} \bar \psi &=& - i e \bar \psi \bar C,
\qquad s_{ab} {\cal B} = 0, \quad  s_{ab} B = 0, \quad
\;s_{ab} E = 0, \quad s_{ab} (\partial \cdot A) = \Box \bar C, 
\end{array}\eqno(2.3)
$$
$$
\begin{array}{lcl}
s_{d} A_{\mu} &=& - \varepsilon_{\mu\nu} \partial^\nu \bar C, \quad
\quad s_{d} B = 0, \quad s_{d} (\partial \cdot A) = 0, \quad
s_{d} \bar C = 0, \quad s_{d} C = - i {\cal B}, \nonumber\\
s_{d} {\cal B} &=& 0, \qquad s_d \psi = - i e \bar C \gamma_5 \psi,
\qquad s_d \bar \psi = + i e \bar \psi \bar C \gamma_5,
\qquad s_{d} E = \Box \bar C, \nonumber\\
s_{ad} A_{\mu} &=& - \varepsilon_{\mu\nu} \partial^\nu  C, \quad
\quad s_{ad} B = 0, \quad s_{ad} (\partial \cdot A) = 0, \quad
s_{ad}  C = 0, \quad s_{ad} \bar C = + i {\cal B}, \nonumber\\
s_{ad} {\cal B} &=& 0, \qquad s_{ad} \psi = - i e C \gamma_5 \psi,
\qquad s_{ad} \bar \psi = + i e \bar \psi  C \gamma_5
\qquad s_{ad} E = \Box  C.
\end{array}\eqno(2.4)
$$
The noteworthy points, at this stage, are (i) under the (anti-)BRST
and (anti-)co-BRST transformations, it is the kinetic energy term 
(more precisely $E$ itself) and the gauge-fixing term
(more accurately $(\partial \cdot A)$ itself) that remain invariant,
respectively. (ii) The electric field $E$ and $(\partial \cdot A)$
owe their origin to the operation of
cohomological operators $d$ and $\delta$ on the one-form $A = dx^\mu A_\mu$,
respectively. (iii) For the (anti-)co-BRST transformations to be the
symmetry transformations for (2.2), there exists the restriction that
$m = 0$ for the Dirac fields. There is no such restriction for the
validity of the (anti-)BRST symmetry transformations. (iv) The 
anticommutator $(s_w = \{ s_{b},  s_{d} \} = \{ s_{ab}, s_{ad} \})$
of the above nilpotent symmetries is a bosonic symmetry 
transformation $s_w$ (with $s_w^2 \neq 0$) respected by the Lagrangian density 
(2.2). These symmetry transformations are
$$
\begin{array}{lcl}
s_{w} A_{\mu} &=& \bigl (\partial_\mu {\cal B} 
+ \varepsilon_{\mu\nu} \partial^\nu B \bigr ), \qquad
s_{w} B = 0, \qquad 
s_{w} C = 0, \nonumber\\
s_{w} (\partial \cdot A) &=&  \Box {\cal B}, \qquad
s_{w} C = 0, \qquad
s_{w} {\cal B} = 0, \qquad 
s_{w} E = - \Box B, \nonumber\\
s_w \psi &=&  i e \bigl (\gamma_5 B - {\cal B}
\bigr )\;\psi,
\quad s_w \bar \psi = + i e \bar \psi \bigl (\gamma_5 B + {\cal B} \bigr).
\end{array}\eqno(2.5)
$$
(v) The operator algebra among the above transformations is exactly
identical to the algebra obeyed by the de Rham cohomological operators
(see, e.g., [36] for details).
(vi) There is a ghost (scale) symmetry transformation under which
only the (anti-)ghost fields transform by a scale factor 
(i.e. $C \to e^{- \Lambda} C, \bar C \to e^{\Lambda} \bar C$)
and the rest of the fields do not transform at all. The
infinitesimal version $s_g$ of this  symmetry transformation is 
$$
\begin{array}{lcl}
s_{g} A_{\mu} = 0, \quad
s_{g} B =  s_g {\cal B} = 0, \quad 
s_{g} \bar C =  + \Lambda \bar C, \quad
s_{g} C = - \Lambda C, \quad
s_g \psi =  s_g \bar \psi = 0,
\end{array}\eqno(2.6)
$$
where $\Lambda$ is a global parameter. (vii) The on-shell nilpotent version 
of the transformations in (2.3) and (2.4) do exist. For this purpose, one
has to get rid of the auxiliary fields $B$ and ${\cal B}$ of the theory
by exploiting the equations of motion (i.e. $B + (\partial\cdot A) = 0,
{\cal B} - E = 0$). These on-shell nilpotent ($ \tilde s_{(a)b}^2 = 0,
\tilde s_{(a)d}^2 = 0$) transformations are
$$
\begin{array}{lcl}
&&\tilde s_{b} A_{\mu} = \partial_{\mu} C, \qquad 
\tilde s_{b} C = 0, \qquad 
\tilde s_{b} \bar C = - i (\partial \cdot A),  \qquad
\tilde s_b E = 0, \nonumber\\
&& \tilde s_b \psi = - i e C \psi, \qquad\;\;
\tilde s_b \bar \psi = - i e \bar \psi C, \;\;\qquad
\tilde s_b (\partial\cdot A) = \Box C, \nonumber\\
&&\tilde s_{ab} A_{\mu} = \partial_{\mu} \bar C, \qquad 
\tilde s_{ab} \bar C = 0, \qquad 
\tilde s_{ab} C = + i (\partial \cdot A),
\qquad \tilde s_{ab} E = 0,  \nonumber\\
&&\tilde s_{ab} \psi = - i e \bar C \psi, \;\;\qquad
\tilde s_{ab} \bar \psi = - i e \bar \psi \bar C, \;\;\;\qquad
\tilde s_{ab} (\partial \cdot A) = \Box \bar C,
\end{array}\eqno(2.7)
$$
$$
\begin{array}{lcl}
&&\tilde s_{d} A_{\mu} = - \varepsilon_{\mu\nu} \partial^\nu \bar C, \qquad
\tilde s_{d} \bar C = 0, \qquad \tilde s_{d} C = - i E, 
\qquad \tilde s_d (\partial \cdot A) = 0, \nonumber\\
&&\tilde s_d \psi = - i e \bar C \gamma_5 \psi,
\qquad \tilde s_d \bar \psi = + i e \bar \psi \bar C \gamma_5, 
\qquad \;\;\; \tilde s_d E = \Box \bar C, \nonumber\\
&&\tilde s_{ad} A_{\mu} = - \varepsilon_{\mu\nu} \partial^\nu  C, \qquad
\tilde s_{ad}  C = 0, \qquad \tilde s_{ad} \bar C = + i E, 
\qquad \tilde s_{ad} (\partial \cdot A) = 0, \nonumber\\
&&s_{ad} \psi = - i e C \gamma_5 \psi,\;\;
\qquad \tilde s_{ad} \bar \psi = + i e \bar \psi  C \gamma_5,
\qquad \;\; \tilde s_{ad} E = \Box C.
\end{array}\eqno(2.8)
$$
(vii) The above symmetry transformations are respected by the
Lagrangian density ${\cal L}_0$ which can be obtained from (2.2)
by the substitution $B = - (\partial \cdot A), {\cal B} = E$, namely;
$$
\begin{array}{lcl}
{\cal L}_{0} =  \frac{1}{2} \;E^2
+ \bar \psi\; (i \gamma^\mu D_\mu - m)\; \psi 
- \frac{1}{2}\; (\partial \cdot A)^2
- i\; \partial_{\mu} \bar C \partial^\mu C.
\end{array} \eqno(2.9)
$$
(viii) It will be noted that, under the on-shell nilpotent symmetry 
transformations, it is only the transformations on the (anti-)ghost fields that
are really affected. The transformations on the gauge- and matter fields remain 
intact. (ix) The above symmetry transformations are generated  by
the local and conserved off-shell as well as on-shell nilpotent
charges $Q_r$ and $\tilde Q_r$, respectively. This
statement can be succinctly expressed in the mathematical form as
$$
\begin{array}{lcl}
s_{r}\; \Omega (x) = - i\; 
\bigl [\; \Omega (x),  Q_r\; \bigr ]_{\pm}, \quad
\tilde s_{r}\; \tilde \Omega (x) = - i\; 
\bigl [\; \tilde \Omega (x),  \tilde Q_r\; \bigr ]_{\pm}, \quad
r = b, ab, d, ad, w, g
\end{array} \eqno(2.10)
$$
where the local generic field
$\Omega = A_\mu, C, \bar C, \psi, \bar \psi, B, 
{\cal B}$ and the $(+)-$ signs, 
as the subscripts on the (anti-)commutator $[\;, \;]_{\pm}$, 
stand for $\Omega$ being (fermionic)bosonic in nature. For the on-shell
nilpotent symmetry transformations in (2.7) and (2.8):
$\tilde \Omega = A_\mu, C, \bar C, \psi, \bar \psi$.

It is worthwhile to mention a set of specific discrete symmetry transformations
in this interacting gauge theory which does correspond to the Hodge duality
$(*)$ operation of the differential geometry. For the off-shell
nilpotent (anti-)BRST invariant Lagrangian density 
(2.2), it can be seen that the following discrete transformations [36]
$$
\begin{array}{lcl}
&&C \to \pm i \gamma_5 \bar C, \quad \psi \to  \psi, \quad
\bar \psi \to \bar \psi, \qquad A_0 \to \pm i \gamma_5 A_1, 
\qquad A_1 \to \pm i \gamma_5 A_0,
\nonumber\\ 
&& \bar C \to  \pm i \gamma_5 C, \qquad e \to \mp i e, \qquad
{\cal B} \to \mp i \gamma_5 B, \qquad B \to \mp i \gamma_5 {\cal B},
\end{array} \eqno(2.11)
$$
correspond to a symmetry transformation as the above transformations
leave the Lagrangian density (2.2) invariant
\footnote{In the matrix notations, the discrete transformations
$A_0 \to \pm i \gamma_5 A_1$, etc., correspond to $A_0 \to \pm i A_1$
and/or  $A_0 \mp i A_1$. Except for the exchange of signs, these 
transformations are identical. In the rest of the transformations (2.11),
$\gamma_5$ has been
 taken into account appropriately to take care of all these sign
flips.}. It can be readily checked that the dual(co-)BRST transformations
in (2.4) can be derived from their counterpart BRST transformations in (2.3)
by exploiting the symmetry transformations listed in (2.11). This claim holds
good for the anti-BRST and anti-co-BRST transformations as well. It is
gratifying to note that, for the generic field $\Omega$, the following
relationship
$$
\begin{array}{lcl} 
s_{(a)d}\; \Omega = \;\pm \; *\; s_{(a)b}\;*\; \Omega,
\end{array} \eqno(2.12)
$$
is sacrosanct. Here the generic field $\Omega = A_0, A_1, B, {\cal B},
C, \bar C, \psi, \bar\psi$ stands for the basic fields of the
Lagrangian density (2.2) and the $*$ operation
corresponds to the transformations (2.11). The $(+)-$ signs in the above 
are dictated by 
a couple of successive transformations (2.11) applied on $\Omega$, as is
required by a consistent duality invariant theory 
(see, e.g., [47])
$$
\begin{array}{lcl} 
*\; (\;*\; \Omega\;) =\; \pm \; \Omega.
\end{array} \eqno(2.13)
$$
It can be readily checked that $(+)$ sign in the above corresponds to
$\Omega$ being $\psi$ and $\bar\psi$ and $(-)$ sign stands for the rest of 
the basic fields of the theory. Thus, we note that the relationship
between the operations $s_d$ and $s_b$ on the generic field $\Omega$
is exactly same as the relationship between the co-exterior derivative
$\delta$ and the exterior derivative $d$ acting on a given differential
form. It is worthwhile to emphasize that the conserved charges
$Q_{(a)b}, Q_{(a)d}, W, Q_g$, that generate the above continuous 
symmetry transformations, undergo the following change under (2.11)
(see, e.g., [36] for details)
$$
\begin{array}{lcl} 
Q_{(a)b} \to\; Q_{(a)d}, \qquad Q_g \to \; Q_g, \qquad
Q_{(a)d} \to \; Q_{(a)b}, \qquad W \to  \;W.
\end{array} \eqno(2.14)
$$
This shows that all the algebraic relations among these charges remain
unchanged under (2.11). The above transformations in (2.14) should be 
contrasted with the corresponding transformations that exist for
BRST invariant four $(3 + 1)$-dimensional 2-form free Abelian gauge 
theory [43,44]
where it has been shown that $Q_{(a)b} \to Q_{(a)d}, 
Q_{(a)d} \to - Q_{(a)b}, W \to - W, Q_g \to - Q_g$ under the analogue
of the discrete transformations corresponding to the 
Hodge duality $(*)$ operation
of differential geometry. This difference in 2D and 4D are
consistent with the key notions connected with the basic idea of the duality
(see, e.g., [47]).\\

\noindent
{\bf 3 Symmetries for Gauge- and (Anti-)ghost Fields:
Usual Superfield Formalism}\\

\noindent
We begin here with a four ($2 + 2$)-dimensional supermanifold
parametrized by the superspace coordinates $Z^M = (x^\mu, \theta, \bar \theta)$
where $x^\mu\; (\mu = 0, 1)$ are a couple of
even (bosonic) spacetime coordinates
and $\theta$ and $\bar \theta$ are the two odd (Grassmannian) coordinates
(with $\theta^2 = \bar \theta^2 = 0, 
\theta \bar \theta + \bar \theta \theta = 0)$. On this supermanifold, one can
define a super 1-form $\tilde A = d Z^M \tilde A_M$
with the supervector superfield $\tilde A_M$ (i.e.
$\tilde A_M = ( B_{\mu} (x, \theta, \bar \theta), 
\;\Phi (x, \theta, \bar \theta),
\;\bar \Phi (x, \theta, \bar \theta))$.
Here $B_\mu, \Phi, \bar \Phi$ are the component
multiplet superfields of the supervector
superfield $\tilde A_M$ [14]. The general
superfields $(B_{\mu}, \Phi, \bar \Phi)(x,\theta,\bar\theta) $ 
can be expanded in terms
of the basic fields $(A_\mu, C, \bar C) (x)$ and  auxiliary fields
($B, {\cal B}) (x)$ of  (2.2) and some extra secondary fields as 
$$
\begin{array}{lcl}
B_{\mu} (x, \theta, \bar \theta) &=& A_{\mu} (x) 
+ \theta\; \bar R_{\mu} (x) + \bar \theta\; R_{\mu} (x) 
+ i \;\theta \;\bar \theta S_{\mu} (x), \nonumber\\
\Phi (x, \theta, \bar \theta) &=& C (x) 
+ i\; \theta \bar B (x)
- i \;\bar \theta\; {\cal B} (x) 
+ i\; \theta\; \bar \theta \;s (x), \nonumber\\
\bar \Phi (x, \theta, \bar \theta) &=& \bar C (x) 
- i \;\theta\;\bar {\cal B} (x) + i\; \bar \theta \;B (x) 
+ i \;\theta \;\bar \theta \;\bar s (x).
\end{array} \eqno(3.1)
$$
It is straightforward to note that the local 
fields $ R_{\mu} (x), \bar R_{\mu} (x),
C (x), \bar C (x), s (x), \bar s (x)$ are the fermionic (anti-commuting) 
in nature and the bosonic (commuting) local fields in (3.1)
are: $A_{\mu} (x), S_{\mu} (x), {\cal B} (x), \bar {\cal B} (x),
B (x), \bar B (x)$. It is obvious
that, in the above expansion, the bosonic-
 and fermionic degrees of freedom do match. This requirement is essential
for the sanctity of any arbitrary supersymmetric field theory in the 
superfield formulation. In fact, all the secondary fields will be expressed 
in terms of basic fields due to the restrictions emerging from the application 
of horizontality condition (i.e. $\tilde F = F$), namely;
$$
\begin{array}{lcl} 
\tilde F =  \frac{1}{2}\; (d Z^M \wedge d Z^N)\;
\tilde F_{MN} \equiv \tilde d \tilde A  =
d A \equiv \frac{1}{2} (dx^\mu \wedge dx^\nu)\; F_{\mu\nu} = F,
\end{array} \eqno(3.2)
$$
where the super exterior derivative $\tilde d$ and 
the connection super one-form $\tilde A$ are defined as
$$
\begin{array}{lcl}
\tilde d &=& \;d Z^M \;\partial_{M} = d x^\mu\; \partial_\mu\;
+ \;d \theta \;\partial_{\theta}\; + \;d \bar \theta \;\partial_{\bar \theta},
\nonumber\\
\tilde A &=& d Z^M\; \tilde A_{M} = d x^\mu \;B_{\mu} (x , \theta, \bar \theta)
+ d \theta\; \bar \Phi (x, \theta, \bar \theta) + d \bar \theta\;
\Phi ( x, \theta, \bar \theta).
\end{array}\eqno(3.3)
$$
In physical language, the requirement (3.2) implies that the 
gauge invariant physical field
$E$, derived from the curvature term $F_{\mu\nu}$, does not get any
contribution from the Grassmannian variables. In other words, the
physical electric field $E$ for 2D QED remains intact in the
superfield formulation. Mathematically, the condition (3.2) implies
the ``flatness'' of all the components of the
super curvature (2-form) tensor $\tilde F_{MN}$ that are directed along the 
 $\theta$ and/or $\bar \theta$ directions of the supermanifold. To clearly
see it, first we expand $\tilde d \tilde A$ as
$$
\begin{array}{lcl}
\tilde d \tilde A &=& (d x^\mu \wedge d x^\nu)\;
(\partial_{\mu} B_\nu) - (d \theta \wedge d \theta)\; (\partial_{\theta}
\bar \Phi) + (d x^\mu \wedge d \bar \theta)
(\partial_{\mu} \Phi - \partial_{\bar \theta} B_{\mu}) \nonumber\\
&-& (d \theta \wedge d \bar \theta) (\partial_{\theta} \Phi 
+ \partial_{\bar \theta} \bar \Phi) 
+ (d x^\mu \wedge d \theta) (\partial_{\mu} \bar \Phi - \partial_{\theta}
B_{\mu}) - (d \bar \theta \wedge d \bar \theta)
(\partial_{\bar \theta} \Phi). 
\end{array}\eqno(3.4)
$$
Ultimately, the application of soul-flatness (horizontality) condition
($\tilde d \tilde A = d A$) yields [27]
$$
\begin{array}{lcl}
R_{\mu} \;(x) &=& \partial_{\mu}\; C(x), \qquad 
\bar R_{\mu}\; (x) = \partial_{\mu}\;
\bar C (x), \qquad \;s\; (x) = \bar s\; (x) = 0,
\nonumber\\
S_{\mu}\; (x) &=& \partial_{\mu} B\; (x) 
\qquad
B\; (x) + \bar B \;(x) = 0, \qquad 
{\cal B}\; (x)  = \bar {\cal B} (x) = 0.
\end{array} \eqno(3.5)
$$
The insertion of all the above values in the expansion (3.1) leads to
the derivation of the (anti-)BRST symmetries for the 
gauge- and (anti-)ghost fields of the Abelian gauge theory.
In addition, this exercise provides  the physical interpretation for the
(anti-)BRST charges $Q_{(a)b}$ 
as the generators (cf. (2.10)) of translations 
(i.e. $ \mbox{Lim}_{\bar\theta \rightarrow 0} (\partial/\partial \theta),
 \mbox{Lim}_{\theta \rightarrow 0} (\partial/\partial \bar\theta)$)
along the Grassmannian
directions of the supermanifold. Both these observations can be succinctly 
expressed, in a combined way, by re-writing the super expansion (3.1) as
$$
\begin{array}{lcl}
B_{\mu}\; (x, \theta, \bar \theta) &=& A_{\mu} (x) 
+ \;\theta\; (s_{ab} A_{\mu} (x)) 
+ \;\bar \theta\; (s_{b} A_{\mu} (x)) 
+ \;\theta \;\bar \theta \;(s_{b} s_{ab} A_{\mu} (x)), \nonumber\\
\Phi\; (x, \theta, \bar \theta) &=& C (x) \;+ \; \theta\; (s_{ab} C (x))
\;+ \;\bar \theta\; (s_{b} C (x)) 
\;+ \;\theta \;\bar \theta \;(s_{b}\; s_{ab} C (x)), 
 \nonumber\\
\bar \Phi\; (x, \theta, \bar \theta) &=& \bar C (x) 
\;+ \;\theta\;(s_{ab} \bar C (x)) \;+\bar \theta\; (s_{b} \bar C (x))
\;+\;\theta\;\bar \theta \;(s_{b} \;s_{ab} \bar C (x)).
\end{array} \eqno(3.6)
$$
The on-shell nilpotent (anti-)BRST symmetry transformations (2.7) can
also be derived in the framework of the usual superfield formulation [27].
For this purpose, we take the chiral limit (i.e. $\theta \to 0$) of the
general expansion in (3.1) and the general definitions in (3.3), as
$$
\begin{array}{lcl}
&& B_{\mu}^{(c)} (x, \bar \theta) = A_{\mu} (x) 
+ \bar \theta\; R_{\mu} (x), \qquad 
\Phi^{(c)} (x, \bar \theta) = C (x) 
- i \;\bar \theta\; {\cal B} (x), \nonumber\\
&&\bar \Phi^{(c)} (x, \bar \theta) = \bar C (x) 
+ i\; \bar \theta \;B (x), \qquad
\tilde d_{(c)} = d x^\mu\; \partial_\mu\;
+ \;d \bar \theta \;\partial_{\bar \theta},
\nonumber\\
&&\tilde A|_{(c)} = d x^\mu \;B_{\mu}^{(c)} (x , \bar \theta)
+ d \bar \theta\;
\Phi^{(c)} ( x, \bar \theta).
\end{array} \eqno(3.7)
$$
It can be noted that the basic tenets of supersymmetry are satisfied 
here too. The bosonic ($A_\mu, B, {\cal B}$) and fermionic 
$(R_\mu, C, \bar C)$ degrees of freedom do match on the chiral three
($2 + 1)$-dimensional super sub-manifold parametrized by two
even $(x_0, x_1)$ variables and one odd $\bar\theta$ variable. 
The explicit form
of the super curvature 2-form on the above super sub-manifold,
constructed by $\tilde d|_{(c)}$ and $\tilde A|_{(c)}$, is
$$
\begin{array}{lcl}
\tilde d|_{(c)} \tilde A|_{(c)} &=& (d x^\mu \wedge d x^\nu)\;
(\partial_{\mu} B_\nu^{(c)}) 
- (d \bar \theta \wedge d \bar \theta)
(\partial_{\bar \theta} \Phi^{(c)})\nonumber\\ 
&+& (d x^\mu \wedge d \bar \theta)
(\partial_{\mu} \Phi^{(c)} - \partial_{\bar \theta} B_{\mu}^{(c)}). 
\end{array}\eqno(3.8)
$$
The horizontality condition $\tilde d|_{(c)} A|_{(c)} = d A$ leads to
the following results:
$$
\begin{array}{lcl}
R_{\mu} \;(x) &=& \partial_{\mu}\; C(x), \qquad 
{\cal B}\; (x)  = 0.
\end{array} \eqno(3.9)
$$
It is evident that the above 
horizontality condition does not fix the value of the
auxiliary field $B$ in terms of the basic fields of the Lagrangian density
(2.9). However, the equation of motion $B + (\partial \cdot A) = 0$,
emerging from  the Lagrangian density
(2.2), comes to our rescue. The insertions of the above values in the
expansion (3.7) lead to the following 
$$
\begin{array}{lcl}
&& B_{\mu}^{(c)} (x, \bar \theta) = A_{\mu} (x) 
+ \bar \theta\; (\tilde s_b A_\mu (x)), \qquad 
\Phi^{(c)} (x, \bar \theta) = C (x) 
+ \bar \theta\; (\tilde s_b C (x)), \nonumber\\
&&\bar \Phi^{(c)} (x, \bar \theta) = \bar C (x) 
+ \; \bar \theta \;(\tilde s_b \bar C (x)). 
\end{array} \eqno(3.10)
$$
Comparison with (2.10) leads to the geometrical interpretation for the
on-shell nilpotent BRST charge $\tilde Q_b$ as the translational generator
$(\partial/\partial \bar\theta)$, for the above chiral superfields,
along the $\bar\theta$-direction of the $(2 + 1)$-dimensional
chiral super sub-manifold. The process of translation of the chiral superfields
produces the on-shell nilpotent transformations $\tilde s_b$ for the usual
fields $A_\mu (x), C (x), \bar C (x)$ defined on the ordinary manifold. Now,
we discuss the derivation of the on-shell nilpotent anti-BRST symmetry
transformations $\tilde s_{ab}$. For this purpose, we take the anti-chiral
(i.e $\bar\theta \to 0$) limit of the general expansion (3.1) as well
the general definitions (3.3), as
$$
\begin{array}{lcl}
&& B_{\mu}^{(ac)} (x, \theta) = A_{\mu} (x) 
+ \theta\; \bar R_{\mu} (x), \qquad 
\Phi^{(ac)} (x, \theta) = C (x) 
- i \;\theta\; B (x), \nonumber\\
&&\bar \Phi^{(ac)} (x, \theta) = \bar C (x) 
- i\; \theta \;\bar {\cal B} (x), \qquad
\tilde d_{(ac)} = d x^\mu\; \partial_\mu\;
+ \;d \theta \;\partial_{\theta},
\nonumber\\
&&\tilde A|_{(ac)} = d x^\mu \;B_{\mu}^{(ac)} (x , \theta)
+ d \theta\;
\bar \Phi^{(ac)} ( x, \theta).
\end{array} \eqno(3.11)
$$
In the above expansion, it should be noted that (i) the fermionic
$(\bar R_\mu, C, \bar C)$ and bosonic $(A_\mu, B, \bar {\cal B})$
degrees of freedom do match, and (ii) we have
taken into account $B (x) + \bar B (x) = 0$, from our earlier
experience in (3.5), for the expansion of $\Phi^{(ac)}$. 
One can construct the super 2-form curvature
$\tilde d|_{(ac)} \tilde A|_{(ac)}$, from the above definitions,
 as follows
$$
\begin{array}{lcl}
\tilde d|_{(ac)} \tilde A|_{(ac)} &=& (d x^\mu \wedge d x^\nu)\;
(\partial_{\mu} B_\nu^{(ac)}) 
- (d \theta \wedge d \theta)
(\partial_{\theta} \bar \Phi^{(ac)})\nonumber\\ 
&+& (d x^\mu \wedge d \theta)
(\partial_{\mu} \bar \Phi^{(ac)} - \partial_{\theta} B_{\mu}^{(ac)}). 
\end{array}\eqno(3.12)
$$
The imposition of the horizontality condition
$\tilde d|_{(ac)} \tilde A|_{(ac)} = d A$ yields the following results
$$
\begin{array}{lcl}
\bar R_{\mu} \;(x) &=& \partial_{\mu}\; \bar C(x), \qquad 
\bar {\cal B}\; (x)  = 0.
\end{array} \eqno(3.13)
$$
It is clear that the above condition (3.13) does not fix the auxiliary field
$B (x)$ present in the expansion of $\Phi^{(ac)}$ in (3.11). However, the
equation of motion $B + (\partial \cdot A) = 0$ derived from the
Lagrangian density (2.2) turns out to be helpful for our purpose. The
substitutions of the results in (3.13) and $B = - (\partial \cdot A)$
in the expansion (3.11) leads to
$$
\begin{array}{lcl}
&& B_{\mu}^{(ac)} (x, \theta) = A_{\mu} (x) 
+ \theta\; (\tilde s_{ab} A_\mu (x)), \qquad 
\Phi^{(ac)} (x, \theta) = C (x) 
+ \theta\; (\tilde s_{ab} C (x)), \nonumber\\
&&\bar \Phi^{(ac)} (x, \theta) = \bar C (x) 
+ \; \theta \;(\tilde s_{ab} \bar C (x)). 
\end{array} \eqno(3.14)
$$
The above expansion explicitly explains the geometrical origin for the
on-shell nilpotent symmetry transformations $\tilde s_{ab}$ as the 
translation generator $(\partial/\partial\theta)$ along the $\theta$-direction
of an anti-chiral super sub-manifold. The process of translation of the
above anti-chiral superfields induces the internal on-shell nilpotent
anti-BRST transformations on the fields $A_\mu (x), C(x), \bar C(x)$
defined on the ordinary manifold.

To obtain the (anti-)co-BRST transformations on the gauge- and (anti-)ghost
fields, we exploit the dual-horizontality condition
$\tilde \delta \tilde A = \delta A$ on the $(2 + 2)$-dimensional
supermanifold where $\tilde \delta = - \star\; \tilde d \; \star$ is the super co-exterior derivative on the four $(2 + 2)$-dimensional supermanifold
and $\delta = - * d *$ is the co-exterior derivative on the ordinary 2D
manifold. The Hodge duality operations on the supermanifold and ordinary
manifold are denoted by $\star$ and $*$, respectively. The $\star$ operations
on the super differentials $(dZ^M)$ and their wedge products
$(dZ^M \wedge dZ^N)$, etc., defined on the $(2 + 2)$-dimensional
supermanifold, are [48]
$$
\begin{array}{lcl}
&&\star\; (dx^\mu) = \varepsilon^{\mu\nu}\; 
(dx_\nu \wedge d \theta \wedge d\bar\theta), \;\;\;\;\;\qquad\;\;\;
\star\; (d\theta) = \frac{1}{2!}\;\varepsilon^{\mu\nu}\; 
(dx_\mu \wedge dx_\nu \wedge d\bar\theta), \nonumber\\
&&\star\; (d\bar\theta) = \frac{1}{2!}\;\varepsilon^{\mu\nu}\; 
(dx_\mu \wedge dx_\nu \wedge d\theta), \;\;\qquad\;\;
\star\; (dx^\mu \wedge dx^\nu) = \varepsilon^{\mu\nu}\; 
(d \theta \wedge d\bar\theta), \nonumber\\
&&\star\; (dx^\mu \wedge d\theta) = \varepsilon^{\mu\nu}\; 
(d x_\nu \wedge d\bar\theta), \;\;\;\qquad\;\;\;\;\;
\star\; (dx^\mu \wedge d\bar\theta) = \varepsilon^{\mu\nu}\; 
(d x_\nu \wedge d\theta), \nonumber\\
&&\star\; (d\theta \wedge d\theta) = \frac{1}{2!}
s^{\theta\theta}\;\varepsilon^{\mu\nu}\; 
(dx_\mu \wedge d x_\nu), \qquad
\star\; (d\theta \wedge d\bar\theta) = \frac{1}{2!}
\varepsilon^{\mu\nu}\; 
(d x_\mu \wedge d x_\nu), \nonumber\\
&&\star\; (d \bar\theta \wedge d\bar\theta) = \frac{1}{2!}
s^{\bar\theta\bar\theta}\;\varepsilon^{\mu\nu}\; 
(dx_\mu \wedge dx_\nu), \;\;\quad\;\; 
\star\;(dx_\mu \wedge d \theta \wedge d\bar\theta) = \varepsilon_{\mu\nu}
(d x^\nu), \nonumber\\
&&\star\;(dx_\mu \wedge dx_\nu
\wedge d \theta \wedge d\bar\theta) = \varepsilon_{\mu\nu}, 
\;\;\;\qquad\;\;\;\;\;
\star\; (dx_\mu \wedge dx_\nu \wedge d \theta) = 
\varepsilon_{\mu\nu} (d \bar \theta), \nonumber\\
&&\star\; (dx_\mu \wedge dx_\nu \wedge d \bar\theta) = 
\varepsilon_{\mu\nu} (d \theta), \;\;\;\;\;\qquad\;\;\;\;
\star \; (dx_\mu \wedge dx_\nu \wedge d\theta \wedge d\theta)
= \varepsilon_{\mu\nu} s^{\theta\theta},
\end{array} \eqno(3.15)
$$
where $s$'s are the symmetric {\it constant} quantities on the Grassmannian
submanifold of the four ($2 + 2$)-dimensional supermanifold. They are
introduced to take care of the fact that two successive $\star$
operation on any differential (or its wedge products)
should yield the same differential (or its wedge products)
(see, e.g., [48] for details). In the above we have collected
only a few of the $\star$ operations. The other appropriate $\star$
operations can be computed in an analogous manner. With the above inputs,
it can be checked that the superscalar superfield $\tilde \delta \tilde A
= - \star \tilde d  \star \tilde A$, expressed in an explicit form,
 turns out to be
$$
\begin{array}{lcl}
\tilde \delta \tilde A = (\partial \cdot B) 
+ (\partial_{\theta} \bar \Phi + \partial_{\bar\theta} \Phi)
+ s^{\theta\theta}
\;(\partial_{\theta} \Phi) + s^{\bar\theta\bar\theta}\;
(\partial_{\bar\theta} \bar \Phi). 
\end{array} \eqno(3.16)
$$
It is to be noted that in the above computation of $\tilde d \star \tilde A$,
we have dropped all the terms that contain (i) more than two differentials
of spacetime in the wedge products, and (ii) more than two differentials
of Grassmann variables in the wedge products. Having done that, we have applied
another $\star$ operation on it
(i.e. $- \star \tilde d \star \tilde A
= \tilde \delta \tilde A$). Ultimately, the dual-horizontality
restriction $\tilde \delta \tilde A = \delta A$ produces the following
restrictions 
$$
\begin{array}{lcl}
\partial_\theta \Phi = 0, \qquad \partial_{\bar\theta} \bar\Phi = 0,
\qquad (\partial \cdot B) + \partial_\theta \bar \Phi
+ \partial_{\bar\theta} \Phi = (\partial \cdot A),
\end{array} \eqno(3.17)
$$
where, as is evident, the r.h.s. of the last entry in the above equation
is due to $\delta A = (\partial \cdot A)$. Exploiting the super 
expansions of (3.1), we obtain
$$
\begin{array}{lcl}
&&(\partial \cdot R) (x) = (\partial \cdot \bar R) (x) 
= (\partial \cdot S) (x) = 0,
\qquad s\; (x) = \bar s \; (x) = 0, \nonumber\\
&& B\; (x) = 0, \quad \bar B \;(x) = 0, \;\;\;\qquad\;\;\;
{\cal B}\; (x) + \bar {\cal B}\; (x) = 0.
\end{array} \eqno(3.18)
$$
It is clear from the above 
that we cannot get a {\it unique} solution for $R_\mu, \bar R_\mu$
and $S_\mu$ in terms of the basic fields of the Lagrangian density (2.2). This
is why there are non-local and non-covariant
solutions for these in the case of QED in 4D.
A detailed discussion on this issue can be found in
our recent work [48]. It is interesting, however, to point out that for 
2D QED, we have a
local and covariant solution for the above restrictions, as
$$
\begin{array}{lcl}
R_\mu = - \varepsilon_{\mu\nu} \partial^\nu \bar C, 
\qquad \bar R_\mu = - \varepsilon_{\mu\nu} \partial^\nu C, \qquad
S_\mu = + \varepsilon_{\mu\nu} \partial^\nu {\cal B}.
\end{array} \eqno(3.19)
$$
With the above insertions, it can be easily checked that the expansion (3.1)
becomes
$$
\begin{array}{lcl}
B_{\mu}\; (x, \theta, \bar \theta) &=& A_{\mu} (x) 
+ \;\theta\; (s_{ad} A_{\mu} (x)) 
+ \;\bar \theta\; (s_{d} A_{\mu} (x)) 
+ \;\theta \;\bar \theta \;(s_{d} s_{ad} A_{\mu} (x)), \nonumber\\
\Phi\; (x, \theta, \bar \theta) &=& C (x) \;+ \; \theta\; (s_{ad} C (x))
\;+ \;\bar \theta\; (s_{d} C (x)) 
\;+ \;\theta \;\bar \theta \;(s_{d}\; s_{ad} C (x)), 
 \nonumber\\
\bar \Phi\; (x, \theta, \bar \theta) &=& \bar C (x) 
\;+ \;\theta\;(s_{ad} \bar C (x)) \;+\bar \theta\; (s_{d} \bar C (x))
\;+\;\theta\;\bar \theta \;(s_{d} \;s_{ad} \bar C (x)).
\end{array} \eqno(3.20)
$$
Thus, the geometrical interpretation for the generators $Q_{(a)d}$ of
the (anti-)co-BRST symmetries is identical to that of the (anti-)BRST
charges $Q_{(a)b}$. However, there is a clear-cut distinction between
$Q_{(a)d}$ and $Q_{(a)b}$ when the transformations on the (anti-)ghost
fields are considered. For instance, the BRST charge $Q_b$ generates
a symmetry transformation such that the  superfield 
$\bar\Phi (x,\theta,\bar\theta)$ becomes {\it anti-chiral} and the 
superfield  $\Phi (x,\theta,\bar\theta)$ becomes an 
ordinary local field $C(x)$. 
In contrast, the co-BRST charge $Q_d$ generates
a symmetry transformation under which just the 
{\it opposite} of the above happens. Similarly, the distinction between
$Q_{ab}$ and $Q_{ad}$ can be argued where one of the above superfields
becomes {\it chiral}.

Let us dwell a bit on the derivations of the on-shell nilpotent 
(anti-)co-BRST symmetry 
transformations $\tilde s_{(a)d}$, listed in  (2.8), by exploiting the
superfield formulation. For this purpose, we focus on the expansions
and definitions, listed in (3.7), for the derivation of the
BRST transformations $\tilde s_b$. With these inputs from (3.7), 
one can compute the following 
$$
\begin{array}{lcl}
\tilde \delta|_{(c)} \tilde A|_{(c)} = (\partial \cdot B^{(c)}) 
+ s^{\bar\theta \bar\theta}
\;(\partial_{\bar \theta} \bar\Phi^{(c)}),
\end{array} \eqno(3.21)
$$
where $\tilde \delta|_{(c)} = - \star\; \tilde d|_{(c)}|\; \star$. It will
be noted that the expression in (3.21) is the chiral limit 
($\theta \to 0$) of the most general expression in (3.16). The 
dual-horizontality condition $\tilde \delta|_{(c)} \tilde A|_{(c)} = \delta A$
on the three $(2 + 1)$-dimensional chiral super sub-manifold leads to
the following conditions on the chiral superfields
$$
\begin{array}{lcl}
 \partial_{\bar\theta} \bar\Phi^{(c)} = 0,
\qquad (\partial \cdot B^{(c)}) = (\partial \cdot A),
\end{array} \eqno(3.22)
$$
which imply a couple of restrictions on the component fields as listed below
$$
\begin{array}{lcl}
(\partial \cdot R) (x) = 0, \qquad\;\;\;
 B\; (x) = 0.
\end{array} \eqno(3.23)
$$
It is evident from the above that there is no {\it unique} solution
to the condition $(\partial \cdot R) = 0$. One can have a non-local 
and non-covariant solution to the above as has been derived for the QED in 
any arbitrary dimensions [37,38]. However,
it is obvious that for the 2D QED, there exists a local and covariant solution
because  $R_\mu$ can be chosen to be: $R_\mu = - \varepsilon_{\mu\nu} 
\partial^\nu \bar C$. It should be noted here that the auxiliary field
${\cal B} (x)$ is not fixed by the above dual-horizontality condition,
in terms of the basic fields of the Lagrangian density (2.1). The equation
of motion $ E - {\cal B} = 0$ from the Lagrangian density (2.2), however, turns
out to be useful for our purpose. The insertions of the above values in
the expansion (3.7) leads to
$$
\begin{array}{lcl}
B_{\mu}^{(c)}\; (x, \bar \theta) &=& A_{\mu} (x) 
+ \;\bar \theta\; (\tilde s_{d} A_{\mu} (x)), \qquad 
\Phi^{(c)}\; (x, \bar \theta) = C (x) \;
\;+ \;\bar \theta\; (\tilde s_{d} C (x)), 
 \nonumber\\
\bar \Phi^{(c)}\; (x, \bar \theta) &=& \bar C (x) 
+\bar \theta\; (\tilde s_{d} \bar C (x)).
\end{array} \eqno(3.24)
$$
To derive the on-shell nilpotent anti-BRST symmetry transformations
of (2.8), we take the anti-chiral limit of the most general expansion
(3.1) and most general definitions in (3.3). These are listed in (3.11).
With these inputs from (3.11), we can define the analogue of (3.21) in terms
of the  anti-chiral super co-exterior derivative
$\tilde \delta|_{(ac)}$ and connection $\tilde A|_{(ac)}$, as
$$
\begin{array}{lcl}
\tilde \delta|_{(ac)} \tilde A|_{(ac)} = (\partial \cdot B^{(ac)}) 
+ s^{\theta \theta}
\;(\partial_{\theta} \Phi^{(ac)}),
\end{array} \eqno(3.25)
$$
where $\tilde \delta_{(ac)} = - \star\; \tilde d|_{(ac)}\; \star$.
The dual-horizontality condition $\tilde \delta|_{(ac)} \tilde A|_{(ac)}
= \delta A$ implies 
$$
\begin{array}{lcl}
 \partial_{\theta} \Phi^{(ac)} = 0 \Rightarrow  B (x) = 0,
\qquad (\partial \cdot B^{(ac)}) = (\partial \cdot A) \Rightarrow
(\partial \cdot \bar R) = 0.
\end{array} \eqno(3.26.)
$$
At this juncture, there are two comments in order. First, the
local and covariant solution for the restriction $(\partial \cdot \bar R) = 0$
exists if we take into account $\bar R_\mu (x) = - \varepsilon_{\mu\nu}
\partial^\nu C$. Second, the above dual-horizontality condition does
not fix the value of the auxiliary field ${\cal B} (x)$. However,
from the equation of motion from the Lagrangian density (2.2), it is clear
that ${\cal B} = E$. The substitution of these values 
(i.e. $ B = 0, {\cal B} = E, \bar R_\mu = - \varepsilon_{\mu\nu}
\partial^\nu C$) leads to the
following expansion for the anti-chiral superfields of (3.11):
$$
\begin{array}{lcl}
B_{\mu}^{(ac)}\; (x, \theta) &=& A_{\mu} (x) 
+ \theta\; (\tilde s_{ad} A_{\mu} (x)), \qquad 
\Phi^{(ac)}\; (x, \theta) = C (x) \;
\;+ \;\bar \theta\; (\tilde s_{ad} C (x)), 
 \nonumber\\
\bar \Phi^{(c)}\; (x, \theta) &=& \bar C (x) 
+\bar \theta\; (\tilde s_{ad} \bar C (x)).
\end{array} \eqno(3.27)
$$
The expansions in (3.27) and (3.24) provide the geometrical
interpretation for the on-shell nilpotent (anti-)co-BRST charges
as the translation generators ($(\partial/\partial\theta)
(\partial/\partial\bar\theta)$)
along the Grassmannian $(\theta)\bar\theta$-directions of the 
three $(2 + 1)$-dimensional (anti-)chiral
super sub-manifolds parametrized by $x^\mu$ and $(\theta)\bar\theta$.
For this interpretation, it is essential to exploit the general
expression for the on-shell nilpotent 
symmetry transformations given in (2.10). To be more precise, the
translations of the (anti-)chiral superfields along the
$(\theta)\bar\theta$-directions of the (anti-)chiral super
sub-manifold generates the internal symmetry transformations
$\tilde s_{(a)d}$ on the usual fields $A_\mu (x), C (x), \bar C (x)$.\\

\noindent
{\bf 4 Nilpotent Symmetries for  Dirac fields:  Augmented Superfield 
Formalism}\\

\noindent
We have exploited the (dual-)horizontality conditions 
(i.e. $\tilde \delta \tilde A = \delta A, \tilde d \tilde A = d A$) on the 
most general 
$(2 + 2)$-dimensional (super)manifolds to derive the (anti-)co-BRST-
and (anti-)BRST symmetry transformations for the gauge- and (anti-)ghost 
fields of the {\it interacting} gauge theory in the previous Section.
It is evident that the horizontality- and dual-horizontality conditions
(which physically
owe their origin to the gauge invariance of the electric field
$E$ and the dual-gauge invariance of the gauge-fixing term, respectively)
do {\it not} shed any light on the nilpotent symmetry transformations
that exist for the matter fields $(\psi, \bar\psi)$ of the interacting theory.
However, we do
know that there exist an {\it additional} gauge invariant quantity
$J_\mu = \bar\psi \gamma_\mu \psi$ and an
{\it additional} $\gamma_5$-chiral (i.e. $ m = 0$) gauge invariant quantity
$J_\mu^{(5)} = \bar\psi \gamma_\mu \gamma_5 \psi$ in the theory. These
invariant quantities
are constructed by the matter fields of the interacting theory (and are
not explicitly written in  terms of the gauge field $A_\mu$
and any of the de Rham cohomological operators). To establish
that the invariance of these currents on the (super)manifolds, leads to 
the derivation of the nilpotent symmetry transformations on the matter fields,
we define the superfields $(\Psi, \bar\Psi)(x, \theta,\bar\theta)$)
corresponding to the ordinary Dirac fields $(\psi, \bar\psi)(x)$. These can
be expanded along all the four independent directions (i.e.
$\hat {\bf 1}, \theta, \bar\theta, \theta\bar\theta$) of the
most general $(2 + 2)$-dimensional supermanifold as [31-33]
$$
\begin{array}{lcl} 
 \Psi (x, \theta, \bar\theta) &=& \psi (x)
+ i \;\theta\; \bar b_1 (x) + i \;\bar \theta \; b_2 (x) 
+ i \;\theta \;\bar \theta \;f (x),
\nonumber\\
\bar \Psi (x, \theta, \bar\theta) &=& \bar \psi (x)
+ i\; \theta \;\bar b_2 (x) + i \;\bar \theta \; b_1 (x) 
+ i\; \theta \;\bar \theta \;\bar f (x).
\end{array} \eqno(4.1)
$$
It is obvious that, in the limit 
$(\theta, \bar\theta) \rightarrow 0$,
we get back the Dirac fields $(\psi, \bar\psi)$ of the 
Lagrangian density (2.1). Furthermore, the number of
bosonic fields ($b_1, \bar b_1, b_2, \bar b_2)$ match with the fermionic
fields $(\psi, \bar \psi, f, \bar f)$ so that the above expansion is consistent
with the basic tenets of supersymmetry for the present superfield
formulation to be correct.

For the sake of one of the the simplest ways to derive the nilpotent BRST
symmetry transformations $s_{b}$ for the
matter fields, we take the chiral limit 
(i.e. $\theta \to 0$) of the expansion (4.1) as illustrated below
$$
\begin{array}{lcl} 
 \Psi^{(c)} (x,  \bar\theta) &=& \psi (x)
+  i \;\bar \theta \; b_2 (x), \qquad 
\bar \Psi^{(c)} (x,  \bar\theta) = \bar \psi (x)
+ i \;\bar \theta \; b_1 (x). 
\end{array} \eqno(4.2)
$$
It can be readily seen that the fermionic
$(\psi, \bar\psi)$ and bosonic $(b_1, b_2)$ degrees of freedom do match
on the chiral supermanifold parameterized by the  bosonic variable $x^\mu$ 
and one
Grassmannian variable $\bar\theta$. From the above chiral expansion,
we can construct the chiral supercurrent $\tilde J_\mu^{(c)} (x,\bar\theta)$ 
and expand it along $\bar\theta$-direction as
$$
\begin{array}{lcl} 
\tilde J_\mu^{(c)} (x, \bar\theta) = \bar \Psi^{(c)} (x, \bar\theta)
\;\gamma_\mu \;\Psi^{(c)} (x, \bar\theta)
= J_\mu (x) 
+ i\;\bar \theta\; \bigl (b_1 \gamma_\mu \psi - \bar \psi \gamma_\mu b_2
\bigr ). 
\end{array} \eqno(4.3)
$$
Requirement of the invariance of the (super)currents on the
(super)manifolds implies 
$$
\begin{array}{lcl} 
\tilde J_\mu^{(c)} (x, \bar\theta) 
= J_\mu (x)\;\;\; \Rightarrow \;\; b_1\;\gamma_\mu\; \psi
= \bar\psi \;\gamma_\mu\; b_2.
\end{array} \eqno(4.4)
$$
The above equality is satisfied if the bosonic components $b_1$ and $b_2$
are proportional to the fermionic fields $\bar\psi$ and $\psi$, respectively.
To make the latter pair $\bar\psi$ and $\psi$ bosonic in nature, we have to
bring in the ghost fields of the theory so that $b_1 \sim \bar\psi C,
b_2 \sim C \psi$. In conformity with the transformations in (2.3), we take
$b_1 = - e \bar\psi C, b_2 = - e C \psi$. Inserting these values in (4.3),
we obtain, {\it vis-{\` a}-vis} (2.3), the following expansion
$$
\begin{array}{lcl} 
 \Psi^{(c)} (x,  \bar\theta) &=& \psi (x)
+   \;\bar \theta \; (s_b \psi (x)), \qquad 
\bar \Psi^{(c)} (x, \bar\theta) = \bar \psi (x)
+  \;\bar \theta \; (s_b \bar \psi (x)). 
\end{array} \eqno(4.5)
$$
This clearly establishes the geometrical interpretation for the nilpotent
BRST charge $Q_{b}$ (and the corresponding transformation $s_b$) as the
translation generator $(\partial/\partial\bar\theta)$ for the chiral
superfields $\Psi^{(c)} (x,\bar\theta)$ and 
$\bar\Psi^{(c)} (x,\bar\theta)$ along the 
$\bar\theta$-direction of the $(2 + 1)$-dimensional chiral super
sub-manifold (parametrized by $x^\mu$ and $\bar\theta$). It will be noted that
the restriction in (4.4) is {\it not} put  by hand. Rather, it is the
inherent property of the theory itself which is quite natural. 
To justify this statement, it can be seen that (4.3) can be re-expressed as
$$
\begin{array}{lcl} 
\tilde J_\mu^{(c)} (x, \bar\theta) = \bar \Psi^{(c)} (x, \bar\theta)
\;\gamma_\mu \;\Psi^{(c)} (x, \bar\theta)
= J_\mu (x) 
+ \;\bar \theta\; \bigl (s_b J_\mu (x) \bigr ). 
\end{array} \eqno(4.6)
$$
However, it can be checked, using the explicit nilpotent BRST transformations
in (2.3), that $s_b J_\mu (x) = 0$. Thus, the restriction
$\tilde J_\mu^{(c)} (x,\bar\theta) = J_\mu (x)$ is {\it not} imposed by
hand from outside. Rather, it is a natural restriction on the supermanifold
which preserves (i) the geometrical interpretation for the $Q_b$ and $s_b$
as the translation generator $(\partial/\partial\bar\theta)$, and
(ii) the nilpotency of $s_b^2 = 0$ (and $Q_b^2 = 0$) as a couple of successive
translations (i.e. $(\partial/\partial\bar\theta)^2 = 0$)
along the $\bar\theta$-direction of the supermanifold. This pair of
geometrical properties (i) and (ii) is exactly same as in the previous
Section where the nilpotent symmetries for the gauge- and (anti-)ghost fields
were derived. Furthermore, it should be noted that
this chiral superfield formulation does {\it not} shed any light on the
anticommutativity $s_b s_{ab} + s_{ab} s_b = 0$ of the (anti-)BRST
transformations $s_{(a)b}$ and corresponding charges $Q_b Q_{ab} 
+ Q_{ab} Q_b = 0$.

One of the easiest ways to derive the nilpotent 
$(s_{ab}^2 = 0)$ anti-BRST transformations
$s_{ab}$ for the matter fields $\psi$ and $\bar\psi$ 
of the Lagrangian density (2.2) is to begin with the
anti-chiral limit ($\bar\theta \to 0$) of the expansion in (4.1) as listed
below
$$
\begin{array}{lcl} 
 \Psi^{(ac)} (x,  \theta) &=& \psi (x)
+  i \;\theta \; \bar b_1 (x), \qquad 
\bar \Psi^{(ac)} (x, \theta, \theta) = \bar \psi (x)
+ i \;\theta \; \bar b_2 (x). 
\end{array} \eqno(4.7)
$$
Here too, the bosonic degrees of freedom $(\bar b_1, \bar b_2)$ and
fermionic degrees of freedom $(\psi, \bar\psi)$ match as per the requirement
of the above theory to be consistent with the basic ideas of supersymmetry.
One can construct the super anti-chiral matter current 
$$
\begin{array}{lcl} 
\tilde J_\mu^{(ac)} (x, \theta) &=& \bar \Psi^{(ac)} (x, \theta)
\;\gamma_\mu \;\Psi^{(ac)} (x, \theta)
= J_\mu (x) 
+ \;\theta\; \bigl (s_{ab} J_\mu (x) \bigr )\nonumber\\
&\equiv& J_\mu (x) + i \;\theta\; (\bar b_2 \gamma_\mu \psi - \bar \psi
\gamma_\mu \bar b_1).
\end{array} \eqno(4.8)
$$
Exploiting the explicit nilpotent transformations $s_{ab}$ from
(2.3), it can be checked that $s_{ab} J_\mu (x) = 0$. Taking this as an
input, the natural restriction on the super sub-manifold
$\tilde J_\mu^{(ac)} (x,\theta) = J_\mu (x)$ implies the following condition
$$
\begin{array}{lcl} 
\bar b_2 (x) \; \gamma_\mu \;\psi (x) = \bar \psi (x)
\gamma_\mu \bar b_1 (x).
\end{array} \eqno(4.9)
$$
As explained earlier in gory detail, it is clear that the above equality 
can be satisfied if one chooses $\bar b_1 = - e \bar C \psi$ and
$\bar b_2 = - e \bar\psi \bar C$. Substituting these values in the
expansion (4.7), we obtain the following expansion in the language
of transformations $s_{ab}$ of (2.3):
$$
\begin{array}{lcl} 
 \Psi^{(ac)} (x,  \theta) &=& \psi (x)
+   \;\theta \; (s_{ab} \psi  (x)), \qquad 
\bar \Psi^{(ac)} (x, \theta) = \bar \psi (x)
+ i \;\theta \; (s_{ab} \bar  \psi (x)). 
\end{array} \eqno(4.10)
$$
The above equation does establish the geometrical interpretation for the
nilpotent charge $Q_{ab}$ (and the corresponding transformation $s_{ab}$)
as the translation generator $(\partial/\partial\theta)$ for the
anti-chiral superfields $\Psi^{(ac)} (x,\theta)$ and 
$\bar \Psi^{(ac)} (x,\theta)$ along the $\theta$-direction of the 
anti-chiral super sub-manifold. The nilpotency
of the charge $Q_{ab}$ (and the corresponding transformation $s_{ab}$)
is encoded in the two successive translations $(\partial/\partial\theta)^2 
= 0$. However, we are {\it not} able to infer any geometrical interpretation
for the anti-commutativity $s_b s_{ab} + s_{ab} s_b = 0$ of the
(anti-)BRST symmetry transformations (as well as the
corresponding charges).

To dwell a bit on the geometrical origin of the anti-commutativity 
$s_b s_{ab} + s_{ab} s_b = 0$ of the transformations (and their
generators $Q_b Q_{ab} + Q_{ab} Q_b = 0$) in the framework of the
general augmented superfield formulation, we focus on the
most general expansion in (4.1). One can construct the
supercurrent $\tilde J_\mu (x, \theta, \bar\theta)$ from the expansion (4.1)
for the superfields with the following general super expansion
$$
\begin{array}{lcl} 
\tilde J_\mu (x, \theta, \bar\theta) = \bar \Psi (x,\theta,\bar\theta)
\;\gamma_\mu \;\Psi (x, \theta, \bar\theta)
= J_\mu (x) + \theta \; \bar K_\mu (x)
+ \bar \theta\; K_\mu (x) + i \; \theta\; \bar\theta\; L_\mu (x), 
\end{array} \eqno(4.11)
$$
where the components $\bar K_\mu, K_\mu, L_\mu, J_\mu$
can be expressed in terms of the components of the
basic super expansions (4.1), as
$$
\begin{array}{lcl} 
&& \bar K_\mu (x) = i \bigl ( \bar b_2 \gamma_\mu \psi -
\bar \psi \gamma_\mu \bar b_1 \bigr ),
\qquad  K_\mu (x) = i \bigl ( b_1 \gamma_\mu \psi -
\bar \psi \gamma_\mu  b_2 \bigr ), \nonumber\\
&& L_\mu (x) = \bar f \gamma_\mu \psi + \bar \psi \gamma_\mu f
+ i (\bar b_2 \gamma_\mu b_2 - b_1 \gamma_\mu \bar b_1), 
\qquad J_\mu (x) = \bar \psi  \gamma_\mu \psi.
 \end{array} \eqno(4.12)
$$
To be consistent with our earlier observation that the BRST
transformations $(s_{b})$ are equivalent to the translations
(i.e. $\mbox{Lim}_{\theta \rightarrow 0} (\partial/\partial \bar\theta)$)
along the $\bar\theta$-direction
and the anti-BRST $(s_{ab})$ transformations
are equivalent to the translations
(i.e. $\mbox{Lim}_{\bar\theta \rightarrow 0} (\partial/\partial \theta)$)
along the $\theta$-direction of the supermanifold, it is straightforward
to re-express the expansion in (4.11) as 
$$
\begin{array}{lcl} 
\tilde J_\mu (x, \theta, \bar\theta) = J_\mu (x) + \theta \; 
(s_{ab} J_\mu (x)) + \bar \theta\; (s_b J_\mu (x)) 
+ \theta\; \bar\theta\; (s_b s_{ab} J_\mu (x)).
\end{array} \eqno(4.13)
$$
It can be checked explicitly that, under the (anti-)BRST transformations (2.3),
the conserved current $J_\mu (x)$ remains invariant
(i.e. $s_{b} J_\mu (x) = s_{ab} J_\mu (x) = 0$).
This statement, with the help of (4.11) and (4.12),
 can be mathematically expressed as
$$
\begin{array}{lcl} 
b_1 \gamma_\mu \psi
= \bar \psi \gamma_\mu b_2, \qquad
\bar b_2 \gamma_\mu \psi
= \bar \psi \gamma_\mu \bar b_1, \qquad
\bar f \gamma_\mu \psi
+ \bar \psi \gamma_\mu f = i (b_1 \gamma_\mu \bar b_1 - \bar b_2
\gamma_\mu b_2). 
\end{array} \eqno(4.14)
$$
One of the possible solutions of the above restrictions, in terms
of the components  of the basic expansions in (4.1) and the
basic fields of the Lagrangian density (2.2), is
$$
\begin{array}{lcl}
&& b_1 = - e \bar \psi C, \qquad b_2 = - e C \psi,
\qquad \bar b_1 = - e \bar C \psi, \qquad \bar b_2 = - e \bar \psi \bar C,
\nonumber\\
&& f = - i e\; [\; B + e \bar C C\; ]\; \psi,
\qquad \bar f = + i e\; \bar \psi\; [\; B + e C \bar C \;].
\end{array} \eqno(4.15)
$$
The above solutions are the
{\it unique} solutions to all the restrictions in (4.14).
Ultimately, the natural restriction that emerges on the $(2 + 2)$-dimensional
supermanifold is
$\tilde J_\mu (x, \theta, \bar \theta ) = J_\mu (x).$
Physically, this mathematical equation implies that there is
no superspace contribution to the ordinary conserved current $J_\mu (x)$. It is
straightforward to check that the substitution of (4.15) into (4.1) 
leads to the following
$$
\begin{array}{lcl}
\Psi\; (x, \theta, \bar \theta) &=& \psi (x) \;+ \; \theta\; 
(s_{ab}  \psi (x))
\;+ \;\bar \theta\; (s_{b} \psi (x)) 
\;+ \;\theta \;\bar \theta \;(s_{b}\;  s_{ab} \psi (x)), 
 \nonumber\\
\bar \Psi\; (x, \theta, \bar \theta) &=& \bar \psi (x) 
\;+ \;\theta\;(s_{ab} \bar \psi (x)) \;+\bar \theta\; (s_{b} \bar \psi (x))
\;+\;\theta\;\bar \theta \;(s_{b} \; s_{ab} \bar \psi (x)).
\end{array} \eqno(4.16)
$$
This establishes the fact that (i) the nilpotent (anti-)BRST charges $Q_{(a)b}$
are the translations generators 
$(\mbox{ Lim}_{\bar \theta \rightarrow 0}(\partial/\partial \theta))
\mbox{ Lim}_{ \theta \rightarrow 0}(\partial/\partial \bar \theta)$ 
along the $(\theta)\bar\theta$ directions
of the supermanifold. (ii) The property of the nilpotency 
(i.e. $Q_{(a)b}^2 = 0$) is encoded in the
two successive translations along the Grassmannian directions of the
supermanifold (i.e. $(\partial/\partial\theta)^2 = 
(\partial/\partial\bar\theta)^2 = 0$). (iii) The anticommutativity
($Q_b Q_{ab} + Q_{ab} Q_b = 0$) property is encoded in
$(\partial/\partial\bar\theta) (\partial/\partial\theta)
 + (\partial/\partial\theta) (\partial/\partial\bar\theta) = 0$ which
also implies $s_b s_{ab} + s_{ab} s_b = 0$.

Now we shall concentrate on the derivation of the symmetry transformations 
(2.4) on the matter fields  in the framework of 
augmented superfield formulation.
To this end in mind, we construct the super $\gamma_5$-axial-vector current
$\tilde J^{(5)}_\mu (x,\theta,\bar\theta)$ and substitute (4.1) to obtain
$$
\begin{array}{lcl} 
\tilde J^{(5)}_\mu (x, \theta, \bar\theta) &=& \bar \Psi (x,\theta,\bar\theta)
\;\gamma_\mu \gamma_5 \;\Psi (x, \theta, \bar\theta) \nonumber\\
&=& J^{(5)}_\mu (x) + \theta \; \bar K^{(5)}_\mu (x)
+ \bar \theta\; K^{(5)}_\mu (x) + i \; \theta\; \bar\theta\; L^{(5)}_\mu (x), 
\end{array} \eqno(4.17)
$$
where the above components on the r.h.s.
can be expressed, in terms of the basic
components of the expansion in (4.1), as
$$
\begin{array}{lcl} 
&& \bar K^{(5)}_\mu (x) = i\; \bigl ( \;\bar b_2 \gamma_\mu \gamma_5 \psi -
\bar \psi \gamma_\mu \gamma_5 \bar b_1 \;\bigr ),
\qquad  K^{(5)}_\mu (x) = i \;\bigl (\; b_1 \gamma_\mu \gamma_5 \psi -
\bar \psi \gamma_\mu  \gamma_5 b_2 \;\bigr ), \nonumber\\
&& L^{(5)}_\mu (x) = \bar f \gamma_\mu \gamma_5 \psi + \bar \psi \gamma_\mu 
\gamma_5 f
+ i (\bar b_2 \gamma_\mu \gamma_5 b_2 - b_1 \gamma_\mu \gamma_5 \bar b_1), 
\qquad J^{(5)}_\mu (x) = \bar \psi  \gamma_\mu \gamma_5 \psi.
 \end{array} \eqno(4.18)
$$
Invoking the  condition $\tilde J^{(5)}_\mu
(x,\theta, \bar\theta) = J^{(5)}_\mu (x)$, we obtain the following
restrictions on the components of the super expansion in (4.17):
$$
\begin{array}{lcl}
K^{(5)}_\mu (x) = 0, \qquad 
\bar K^{(5)}_\mu (x) = 0, \qquad 
L^{(5)}_\mu (x) = 0. 
\end{array} \eqno(4.19)
$$
Ultimately, as discussed earlier in gory details, these conditions lead to 
$$
\begin{array}{lcl}
&& b_1 = + e \bar \psi \bar C \gamma_5, \qquad 
b_2 = - e \bar C \gamma_5 \psi,
\qquad \bar b_1 = - e C \gamma_5 \psi, 
\qquad \bar b_2 = + e \bar \psi C \gamma_5,
\nonumber\\
&& f = + i e\; [ \;{\cal B} \gamma_5 - e C \bar C \;] \;\psi,
\qquad \bar f = + i e\; \bar \psi\; [\;{\cal B}\gamma_5 +
 e \bar C C \;].
\end{array} \eqno(4.20)
$$
The substitution of the above values in the super expansion in (4.1)
leads to the analogous expansion as in (4.16) with the replacements:
$s_b \rightarrow s_d, \; s_{ab}\rightarrow s_{ad}$. Thus, we obtain
$$
\begin{array}{lcl}
\Psi\; (x, \theta, \bar \theta) &=& \psi (x) \;+ \; \theta\; 
(s_{ad}  \psi (x))
\;+ \;\bar \theta\; (s_{d} \psi (x)) 
\;+ \;\theta \;\bar \theta \;(s_{d}\;  s_{ad} \psi (x)), 
 \nonumber\\
\bar \Psi\; (x, \theta, \bar \theta) &=& \bar \psi (x) 
\;+ \;\theta\;(s_{ad} \bar \psi (x)) \;+\bar \theta\; (s_{d} \bar \psi (x))
\;+\;\theta\;\bar \theta \;(s_{d} \; s_{ad} \bar \psi (x)).
\end{array} \eqno(4.21)
$$
This provides the
 geometrical interpretation for the nilpotent (anti-)co-BRST charges as
the translation generators along the $(\theta)\bar\theta$-directions
of the supermanifold. The geometrical interpretation for the
nilpotency and anticommutativity properties is found to be
exactly identical to the case of
(anti-)BRST charges (as discussed after (4.16)) .

It is very interesting to point out that the (anti-)co-BRST 
transformations of (4.21) can be derived separately and independently as well.
For this purpose, first we take the chiral limit (i.e. $\theta \to 0$)
of the most general expansion in (4.1) so that we can obtain the
expansion in (4.2) for the chiral superfields $\Psi^{(c)} (x,\bar\theta)$
and $\bar\Psi^{(c)} (x,\bar\theta)$. We can construct a $\gamma_5$-chiral
supercurrent in terms of the expansion in (4.2) as
$$
\begin{array}{lcl} 
\tilde J_\mu^{(5c)} (x, \bar\theta) = \bar \Psi^{(c)} (x, \bar\theta)
\;\gamma_\mu \;\gamma_5\;\Psi^{(c)} (x, \bar\theta)
= J_\mu^{(5)} (x) 
+ i\;\bar \theta\; \bigl (b_1 \gamma_\mu \gamma_5\psi 
- \bar \psi \gamma_\mu \gamma_5 b_2
\bigr ). 
\end{array} \eqno(4.22)
$$
It is clear that the above expansion in (4.22) can be re-expressed,
{\it vis-{\`a}-vis} (4.13), as
$$
\begin{array}{lcl} 
\tilde J_\mu^{(5c)} (x, \bar\theta) = \bar \Psi^{(c)} (x, \bar\theta)
\;\gamma_\mu \;\gamma_5\;\Psi^{(c)} (x, \bar\theta)
= J_\mu^{(5)} (x) 
+ \bar \theta\; \bigl (s_d \; J_\mu^{(5)} (x) \bigr ). 
\end{array} \eqno(4.23)
$$
However, it is can be readily checked, using (2.4), that
$s_d J_\mu^{(5)} (x) = 0$. Thus, we obtain a natural restriction on
the (super)manifolds as $\tilde J_\mu^{(5c)} (x,\theta) = J_\mu^{(5)} (x)$.
In the language of the expansion in (4.22), this restriction can be
satisfied by the choices of the components $b_1$ and $b_2$ as
$ b_1 = + e \bar \psi \bar C \gamma_5$ and 
$ b_2 = - e \bar C \gamma_5 \psi$. Insertions of these values into
the chiral expansion in (4.2) leads to the following expansion in the
language of the transformations (2.4)
$$
\begin{array}{lcl}
\Psi^{(c)}\; (x, \bar \theta) = \psi (x) 
\;+ \;\bar \theta\; (s_{d} \psi (x)), \qquad 
\bar \Psi^{(c)}\; (x, \bar \theta) = \bar \psi (x) 
+\bar \theta\; (s_{d} \bar \psi (x)).
\end{array} \eqno(4.24)
$$
Taking the help of our generic expression in (2.10), it is clear that the 
conserved and
nilpotent dual(co-)BRST charge and the corresponding transformation
$s_d$ geometrically imply the translation of the chiral superfields
$\Psi^{(c)} (x, \bar\theta)$ and $\bar\Psi^{(c)} (x,\bar\theta)$ along the
$\bar\theta$-direction of the chiral super sub-manifold parametrized
by even variables $x^\mu$ and odd variable $\bar\theta$. In a similar fashion,
we can obtain the nilpotent anti-co-BRST transformations by taking into
account the anti-chiral limit (i.e. $\bar\theta \rightarrow 0$) of the 
expansion in (4.1) and, thereby, obtain the expansion given in (4.7).
The latter equation allows us to write the $\gamma_5$-anti-chiral supercurrent
in the expanded form as given below
$$
\begin{array}{lcl} 
\tilde J_\mu^{(5ac)} (x, \theta) = \bar \Psi^{(ac)} (x, \theta)
\;\gamma_\mu \;\gamma_5\;\Psi^{(ac)} (x, \theta)
= J_\mu^{(5)} (x) 
+ i\;\theta\; \bigl (\bar b_2 \gamma_\mu \gamma_5\psi 
- \bar \psi \gamma_\mu \gamma_5 \bar b_1
\bigr ). 
\end{array} \eqno(4.25)
$$
On the other hand, the anti-chiral ($ \bar\theta \to 0$) limit of the
expansion in (4.17) leads to the following expansion ({\it vis-{\`a}-vis} 
(4.13)):
$$
\begin{array}{lcl} 
\tilde J^{(5ac)}_\mu (x, \theta, \bar\theta) 
= \bar \Psi^{(ac)} (x,\theta)
\;\gamma_\mu \gamma_5 \;\Psi (x, \theta) 
= J^{(5)}_\mu (x) + \theta \; (s_{ad} J_\mu^{(5)} (x)).
\end{array} \eqno(4.26)
$$
Exploiting the explicit anti-co-BRST transformations of (2.4), it can be 
readily seen
that $s_{sd} J_\mu^{(5)} (x) = 0$. It is obvious that the consistency and
conformity between (4.26) and (4.25) implies that
$\bar b_2 \gamma_\mu \gamma_5\psi 
= \bar \psi \gamma_\mu \gamma_5 \bar b_1$. This restriction can be satisfied
by the choice of $\bar b_1$ and $\bar b_2$ as
$\bar b_1 = - e C \gamma_5 \psi, 
\; \bar b_2 = + e \bar \psi C \gamma_5$. Plugging in these values in (4.7)
leads to the following expansion for the anti-chiral superfields
$$
\begin{array}{lcl}
\Psi^{(ac)} \; (x, \theta) &=& \psi (x) \;+ \; \theta\; 
(s_{ad}  \psi (x)), \qquad
\bar \Psi^{(ac)}\; (x, \theta) = \bar \psi (x) 
\;+ \;\theta\;(s_{ad} \bar \psi (x)).
\end{array} \eqno(4.27)
$$
The above equation provides the geometrical interpretation for the
nilpotent anti-co-BRST charge $Q_{ad}$ and the corresponding 
nilpotent transformation $s_{ad}$ as the translational generator
$(\partial/\partial\theta)$ for the anti-chiral superfields
$\Psi^{(ac)} (x,\theta)$ and $\bar\Psi^{(ac)} (x,\theta)$ along the
$\theta$-direction of the anti-chiral three $(2 + 1)$-dimensional
supermanifold parametrized
by a couple of even variables $x^\mu (\mu = 0, 1)$ and an odd 
variable $\theta$. It is clear that the super translation 
$(\partial/\partial\theta)$ of the above
anti-chiral superfields along the $\theta$-direction of
the supermanifold generates the internal nilpotent symmetry 
transformations $s_{ad}$ on the ordinary fermionic Dirac fields 
$\psi (x)$ and $\bar\psi (x)$ located on the ordinary manifold,
parametrized by $x^\mu$.

To wrap up this Section, we make a {\it couple} of
 general remarks on the discrete symmetry
transformations in (2.11) as well as the
nilpotent (anti-)BRST- and (anti-)co-BRST symmetry transformations
(and their corresponding nilpotent generators). First, it can be checked 
that the supersymmetric counterpart of the discrete transformations
(2.11) are
$$
\begin{array}{lcl}
&&C \to \pm i \gamma_5 \bar C, \quad \psi \to  \psi, \quad
\bar \psi \to \bar \psi, \qquad A_0 \to \pm i \gamma_5 A_1, 
\qquad A_1 \to \pm i \gamma_5 A_0,
\nonumber\\ 
&& \bar C \to  \pm i \gamma_5 C, \qquad e \to \mp i e, \qquad
{\cal B} \to \mp i \gamma_5 B, \qquad B \to \mp i \gamma_5 {\cal B},
\nonumber\\
&& \theta \to \theta, \quad \bar\theta \to \bar\theta, \quad
S_0 \to \pm i \gamma_5 S_1, \quad S_1 \to \pm i \gamma_5 S_0, \quad
b_1 \to b_1, \quad b_2 \to b_2,
\nonumber\\
&& R_0 \to \pm i \gamma_5 R_1, \qquad \bar R_0 \to \pm i \gamma_5 \bar R_1,
\qquad
 R_1 \to \pm i \gamma_5 R_0, \qquad \bar R_1 \to \pm i \gamma_5 \bar R_0,
\nonumber\\
&&
\bar b_{1} \to \bar b_{1}, \qquad \bar b_2 \to \bar b_2, \qquad
f \to f, \qquad \bar f \to \bar f.
\end{array} \eqno(4.28)
$$
It is straightforward to check that, under the above discrete transformations,
the superfields in (3.1) and (4.1) transform in analogous manner as their
ordinary counterparts. These are 
$$
\begin{array}{lcl}
&&B_0 \to \pm i \gamma_5 B_1, \qquad
B_1 \to \pm i \gamma_5 B_0, \qquad \Phi \to \pm i \gamma_5 \bar \Phi,
\nonumber\\
&& \bar \Phi \to \pm i \gamma_5 \Phi, 
\qquad\;\;
\Psi \to \Psi, \qquad\; \; \;
\bar\Psi \to \bar \Psi.
\end{array} \eqno(4.29)
$$
The above transformations are the analogue of the $\star$ operations
defined in (3.15) for the superspace differentials and their wedge
products. However, the operation of the $\star$, listed in the
language of the discrete symmetry transformations in (4.29), is applicable
in the space of superfields. To obtain the analogue of equation (2.12),
we have to express the nilpotent (anti-)BRST and (anti-)co-BRST symmetry
transformations in the language of the superfields. We have not achieved
this goal in our present paper. We hope to come to it later. Second,
the {\it common} geometrical interpretations
for the nilpotent (anti-)BRST- and (anti-)co-BRST charges
can be succinctly expressed in the mathematical form, using 
the general expression (2.10) for the transformations, as 
$$
\begin{array}{lcl}
s_{p} \Omega (x) = \mbox{Lim}_{\theta \rightarrow 0} 
{\displaystyle \frac{\partial}{\partial
\bar\theta}}  \Omega^{(s)} (x,\theta,\bar\theta)
\equiv - i \bigl [ \Omega (x), Q_{p} \bigr ]_{\pm}, \nonumber\\
s_{q} \Omega (x) = \mbox{Lim}_{\bar \theta \rightarrow 0} 
{\displaystyle \frac{\partial}{\partial
\theta}} \Omega^{(s)} (x,\theta,\bar\theta)
\equiv - i \bigl [ \Omega (x), Q_{q} \bigr ]_{\pm}, 
\end{array} \eqno(4.30)
$$
where $p = b, d$,  $\;q = ab, ad$ and $\Omega (x) 
= \psi (x), \bar \psi (x), C (x), \bar C (x), A_\mu (x)$ and
$\Omega^{(s)} (x,\theta,\bar\theta) = \Psi (x,\theta, \bar\theta), 
\bar \Psi (x,\theta,\bar\theta), \Phi (x,\theta,\bar\theta),
\bar \Phi (x,\theta,\bar\theta), B_\mu (x,\theta,\bar\theta)$. The
$(+)-$ signs on the brackets correspond to the (anti-)commutators for
the generic field $\Omega (x)$ being (fermionic)bosonic in nature. 
Thus, it is clear that the following mapping exists
among the symmetry transformations, the conserved charges and 
the translation generators along the Grassmannian directions
$$
\begin{array}{lcl}
&&s_{b} \;\;\leftrightarrow \;\;Q_{b} \;\;\leftrightarrow\;\;
\mbox{Lim}_{\theta \rightarrow 0} {\displaystyle \frac{\partial}{\partial
\bar\theta}}, \qquad
s_{d} \;\;\leftrightarrow \;\;Q_{d} \;\;\leftrightarrow\;\;
\mbox{Lim}_{\theta \rightarrow 0} {\displaystyle \frac{\partial}{\partial
\bar\theta}}, \nonumber\\
&& s_{ad} \;\;\leftrightarrow\;\; Q_{ad} \;\;\leftrightarrow\;\;
\mbox{Lim}_{\bar \theta \rightarrow 0} {\displaystyle \frac{\partial}{\partial
\theta}}, \qquad
s_{ab} \;\;\leftrightarrow \;\;Q_{ab} \;\;\leftrightarrow\;\;
\mbox{Lim}_{\bar\theta \rightarrow 0} {\displaystyle \frac{\partial}{\partial
\theta}}.
\end{array} \eqno(4.31)
$$
It will be noted that in equations 
(4.30) and (4.31), we have taken into account
only the off-shell nilpotent symmetries and their corresponding generators.
However, it is straightforward to express these kind of relations for
the on-shell nilpotent symmetries and their corresponding 
generators as well. \\

\noindent
{\bf 5 Conclusions}\\

\noindent
The central theme of our present investigation
was to provide the geometrical origin and 
interpretation for the off-shell as well as on-shell nilpotent symmetries
(and their corresponding nilpotent generators) in the language of the 
translations along the Grassmannian directions of the supermanifold.
These twin objectives have been achieved by exploiting the augmented
superfield formulation for the case of the 2D QED considered on the general
four $(2 + 2)$-dimensional supermanifold as well as its (anti-)chiral
three $(2 + 1)$-dimensional super sub-manifolds, respectively.
The key roles in our present endeavour are played by the 
(dual-)horizontality conditions, invariance of the matter 
(super)currents and the nilpotent super
de Rham cohomological operators $\tilde d$ and $\tilde\delta 
= - \star \tilde d \star$ (and
their ordinary  counterparts $d$ and
$\delta = - * d *$) on the (super)manifolds. 
In particular, 
the nilpotent cohomological operators are useful
in (i) providing the basis for
the (dual-)horizontality conditions imposed on the supermanifolds,
(ii) providing the derivations of the nilpotent symmetries for the gauge-
and (anti-)ghost fields of the interacting gauge theory (i.e. 2D QED), 
(iii) providing the geometrical interpretations for the nilpotent symmetries
and their corresponding generators, and (iv) providing the origin for the
existence of the kinetic energy and gauge-fixing terms of the
Lagrangian density (cf.(2.1) and (2.2)) of the BRST invariant gauge theory.

As discussed in Section 2, the cohomological operators $(d, \delta, \Delta)$
find their physical interpretation in the language of the symmetry properties
of the BRST invariant Lagrangian density (2.2) for the 2D QED. The invariance
of the kinetic energy term (constructed by $d$ and $A$) is at the heart
of the existence of the (anti-)BRST symmetry transformations. In contrast,
the (anti-)co-BRST symmetry transformations owe their existence to the
invariance of the gauge-fixing term (constructed by $\delta$ and $A$)
of the Lagrangian density (2.2) of 2D QED. The analogue of the
Laplacian operator $\Delta$ is a bosonic symmetry transformation
(equal to the anticommutator of the nilpotent
(anti-)BRST and (anti-)co-BRST symmetry transformations)
under which the ghost term of the Lagrangian density (2.2)
does not transform 
\footnote{ The key role played by $\Delta$, however,  is in the
context of the equation of motion for the gauge field $A_\mu$ for
the case of 2D QED 
which is described by the gauge-fixed Lagrangian density (2.2).}. 
As a consequence, the generators
of the above symmetry transformations $Q_{(a)b}, Q_{(a)d}$ and $W
= \{Q_b, Q_d\} = \{Q_{ad}, Q_{ab} \}$ 
mathematically rely for their existence to the
cohomological operators $d, \delta, \Delta$. In fact, the ghost number
consideration allows us to obtain the following two-to-one mapping
from the set of local and
conserved charges to the set of the de Rham cohomological operators 
$$
\begin{array}{lcl}
(Q_b, Q_{ad}) \rightarrow d, \qquad
(Q_d, Q_{ab}) \rightarrow \delta, \qquad
W = \{Q_b, Q_d \} = \{Q_{ad}, Q_{ab} \} \rightarrow \Delta.
\end{array} \eqno(5.1)
$$
Due to the above equation, any arbitrary state $|\phi>_n$ in the quantum
Hilbert space,
with the ghost number $n$ (i.e. $i Q_g |\phi>_n = n \; |\phi>_n$), can
be decomposed into a harmonic state $|h>_n$
(with $Q_{b} |h>_n = Q_{d} |h>_n = 0$), a BRST exact state
$Q_b |\theta>_{n - 1}$ and a BRST co-exact state $Q_d |\chi>_{n + 1}$
as given below [27,35,36]
$$
\begin{array}{lcl}
|\phi>_n = |h>_n + Q_b |\theta>_{n-1} + Q_d |\chi>_{n + 1}
 = |h>_n + Q_{ad} |\theta>_{n-1} + Q_{ab} |\chi>_{n + 1}.
\end{array} \eqno(5.2)
$$
The above equation is the analogue of the Hodge decomposition theorem
for a differential form (cf. the footnote on page 3) 
(see, e.g., [20-24] for details).

One of the key observations of our present investigation is the consistency
and complementarity between the (dual-)horizontality conditions and
the requirement of the invariance of the matter (super)currents on the 
(super)manifolds. It is obvious that the augmented superfield
formulation is the generalization of the  usual superfield formalism
because (i) the geometrical interpretation
for the nilpotent charges in the language of the translation generators,
along the Grassmannian directions, remains intact, (ii)
the nilpotency of the above conserved charges 
(as well as the cohomological operators) is found to be encoded in 
the property $(\partial/\partial\theta)^2 = 0, (\partial/\partial\bar\theta)^2
= 0$. Geometrically, the above relations imply that two successive
translations of any superfields along {\it either} of the
two Grassmannian directions corresponds to {\it no} translations at all,
(iii) the geometrical interpretation for the anticommutativity of
$s_b s_{ab} + s_{ab} s_b = 0$ and $s_d s_{ad} + s_{ad} s_d = 0$ is
hidden in the relation $(\partial/\partial\theta) (\partial/\partial\bar\theta)
+ (\partial/\partial\bar\theta) (\partial/\partial\theta) = 0$, (iv)
the nilpotent transformations on the gauge- and (anti-)ghost fields
derived from the (dual-)horizontality condition are found to
be consistent with the analogous
 transformations on the matter fields derived due
to the requirements of the invariance of the matter (super)currents on the
(super)manifolds.

The consistency and complementarity between the restrictions due to
(i) the horizontality condition, and (ii) the invariance of the 
conserved matter (super)current on the (super)manifolds can be understood
in the language of the {\it gauge invariance}. It is known that, for
the Abelian gauge theory, the 2-form curvature $F = d A$ remains invariant
under the gauge (or BRST) transformation $A \to A + d C$ which involves
the gauge- and ghost fields. This is why, the horizontality condition
$\tilde d \tilde A = d A$ produces the nilpotent BRST transformations on
the gauge- and ghost fields. In fact, the gauge (or BRST) transformations
on the matter fields in (2.3) owe their origin to the gauge transformation
$A \to A + d C$. Thus, the requirement of the matter conserved (super)currents
$\tilde J_\mu (x,\theta,\bar\theta) = J_\mu (x)$ on the (super)manifolds
yields the nilpotent BRST transformations on the matter fields that are found 
to be consistent with the nilpotent transformations on the gauge-
and (anti-)ghost fields. In the language of the superfield formalism, there
is a key difference between the derivation of the nilpotent symmetries
for the gauge- as well as (anti-)ghost fields and the matter fields. Whereas 
the choice of the (anti-)chiral superfields yields the on-shell nilpotent
symmetries for the gauge- and (anti-)ghost fields, the same choices do
not produce a {\it different} 
kind of symmetry transformations for the matter fields.
In other words, the choices of the (anti-)chiral superfields have no major
effect on the transformations of the matter fields.

It would be very nice and challenging endeavour to extend our present
work to the case of 2D interacting non-Abelian gauge theory where there
is an interaction between the matter fields and the non-Abelian gauge field.
In fact, the (anti-)BRST symmetries for all the fields present in the
4D BRST invariant Lagrangian density of the non-Abelian gauge theory
have already been derived in the augmented superfield formalism [33].
However, the non-local, non-covariant, continuous and
nilpotent (anti-)co-BRST symmetries that exist 
(see, e.g., [37,38,48,49] for details) for the 4D (non-)Abelian
gauge theories have {\it not} 
yet been derived in this augmented superfield
approach. 
These are some of the open problems which are under investigation and our 
results will be reported elsewhere [50].

\end{document}